\documentclass[journal,10pt]{IEEEtran}
\pdfoutput=1
\usepackage{cite}
\usepackage[cmex10]{amsmath}
\usepackage{multirow}
\usepackage{amssymb}
\usepackage{booktabs}
\usepackage{array}
\usepackage{adjustbox}
\usepackage[lofdepth,lotdepth]{subfig}
\usepackage{lipsum}
\usepackage{mathtools}
\usepackage{cuted}
\usepackage{breqn}
\usepackage{graphicx}
\usepackage{epstopdf}
\usepackage{gensymb}
\usepackage[english]{babel}
\usepackage[autostyle]{csquotes}
\usepackage[acronym,toc,shortcuts]{glossaries}
\usepackage{caption}
\usepackage{algorithmicx}
\usepackage[noend]{algpseudocode}
\usepackage{amsmath}
\usepackage{verbatim}
\usepackage[graphicx]{realboxes}
\usepackage{euscript}
\usepackage{color}
\usepackage{url}
\usepackage{cite}
\usepackage{multirow}

\makeglossaries
\newacronym{BER}{BER}{bit-error-rate}
\newacronym{APD}{APD}{avalanche photo detector}
\newacronym{ANN}{ANN}{artificial neural network}
\newacronym{AWG}{AWG}{arbitrary waveform generator}
\newacronym{CC}{CC}{Convolutional Code}
\newacronym{CDF}{CDF}{cumulative distribution function}
\newacronym{CamCom}{CamCom}{optical camera communications}
\newacronym{CFR}{CFR}{channel frequency response}
\newacronym{CLT}{CLT}{Central Limit Theorem}
\newacronym{CIR}{CIR}{channel impulse response}
\newacronym{CMOS}{CMOS}{Complementary Metal Oxide Semiconductor}
\newacronym{CNN}{CNN}{convolutional neural network}
\newacronym{COTS}{COTS}{commercial off-the-shelf}
\newacronym{CSK}{CSK}{Color Shift Keying}
\newacronym{DLoS}{DLoS}{Directed Line of Sight}
\newacronym{DSRC}{DSRC}{Dedicated Short Range Communication}
\newacronym{DS1}{DS1}{dataset 1}
\newacronym{DS2}{DS2}{dataset 2}
\newacronym{DS3}{DS3}{dataset 3}
\newacronym{DRL}{DRL}{day time running light}
\newacronym{FEC}{FEC}{Forward Error Correction}
\newacronym{FF-NN}{FF-NN}{feedforward neural network}
\newacronym{FLP}{FLP}{fast locking pattern}
\newacronym{FoV}{FoV}{field of view}
\newacronym{FPGA}{FPGA}{Field Programmable Gate Array}
\newacronym{GBSM}{GBSM}{geometry based stochastic models}
\newacronym{GPS}{GPS}{global positioning system}
\newacronym{fps}{fps}{frame per second}
\newacronym{IF}{IF}{intermediate frequency}
\newacronym{IM/DD}{IM/DD}{intensity modulation and direct detection}
\newacronym{ISI}{ISI}{intersymbol interference}
\newacronym{ITS}{ITS}{intelligent transportation systems}
\newacronym{KNN}{KNN}{k-nearest neighbor classifier}
\newacronym{LED}{LED}{light emitting diode}
\newacronym{LNA}{LNA}{low noise amplifier}
\newacronym{LoS}{LoS}{line-of-sight}
\newacronym{LTE}{LTE}{Long Term Evolution}
\newacronym{MAC}{MAC}{medium access control}
\newacronym{MAE}{MAE}{mean absolute error}
\newacronym{ML}{ML}{machine learning}
\newacronym{MLP}{MLP}{multilayer perceptron}
\newacronym{mmWave}{mmWave}{millimeter-wave}
\newacronym{MPC}{MPC}{multi path component}
\newacronym{MCS}{MCS}{modulation coding schemes}
\newacronym{MIMO}{MIMO}{multiple input multiple output}
\newacronym{MSE}{MSE}{mean square error}
\newacronym{NLoS}{NLoS}{non-line-of-sight}
\newacronym{NN}{NN}{neural network}
\newacronym{NoR}{NoR}{norm of residuals}
\newacronym{OCC}{OCC}{optical  camera  communications}
\newacronym{OLED}{OLED}{organic light emitting diode}
\newacronym{OOK}{OOK}{on-off keying modulation}
\newacronym{OWC}{OWC}{Optical Wireless Communication}
\newacronym{PCA}{PCA}{Principal Component Analysis}
\newacronym{PD}{PD}{photodetector}
\newacronym{PDP}{PDP}{power delay profile}
\newacronym{PHR}{PHR}{physical header}
\newacronym{PHY}{PHY}{physical layer}
\newacronym{PMT}{PMT}{photomultiplier tube}
\newacronym{PPM}{PPM}{pulse position modulation}
\newacronym{PWM}{PWM}{pulse width modulation}
\newacronym{PSDU}{PSDU}{physical service data unit}
\newacronym{RBF}{RBF}{radial basis function}
\newacronym{RBF-NN}{RBF-NN}{radial basis function neural networks}
\newacronym{RF}{RF}{radio frequency}
\newacronym{R^2}{$R^2$}{Coefficient of Determination (R-Squared)}
\newacronym{RLL}{RLL}{Run-Length Limited}
\newacronym{RMS}{RMS}{root mean square}
\newacronym{RMSE}{RMSE}{root mean square error}
\newacronym{RS}{RS}{Reed Solomon}
\newacronym{RSS}{RSS}{received signal strength}
\newacronym{RU}{RU}{Receiver unit}
\newacronym{SDR}{SDR}{Software-Defined Radio}
\newacronym{SHR}{SHR}{synchronization header}
\newacronym{SNR}{SNR}{signal to noise ratio}
\newacronym{SPAD}{SPAD}{single photon avalanche diode}
\newacronym{SSE}{SSE}{error sum of squares}
\newacronym{TDP}{TDP}{topology dependent pattern}
\newacronym{TU}{TU}{Transmitter Unit}
\newacronym{USRP}{USRP}{Universal Software Radio Peripheral}
\newacronym{VSA}{VSA}{vector signal analyzer}
\newacronym{V2I}{V2I}{vehicular to infrastructure}
\newacronym{V2V}{V2V}{vehicle to vehicle}
\newacronym{V2X}{V2X}{vehicle to everything}
\newacronym{V-VLC}{VVLC}{vehicular visible light communication}
\newacronym{VVLC}{VVLC}{vehicular visible light communication}
\newacronym{VLC}{VLC}{visible light communication}
\newacronym{VNA}{VNA}{vector network analyzer}
\newacronym{VPPM}{VPPM}{variable pulse position modulation}

\begin{document}

\title{Machine Learning Based Channel Modeling for Vehicular Visible Light Communication}

\author{Bugra~Turan,~\IEEEmembership{Student Member,~IEEE}, {Sinem~Coleri,~\IEEEmembership{Senior~Member,~IEEE}}

\thanks{B. Turan and S. Coleri are with the Department of Electrical and Electronics Engineering, Koc University, Sariyer, Istanbul, Turkey. \protect\\
E-mail: bturan14@ku.edu.tr, scoleri@ku.edu.tr}}



\maketitle
\begin{abstract}


Optical Wireless Communication (OWC) propagation channel characterization plays a key role on the design and performance analysis of Vehicular Visible Light Communication (VVLC) systems. Current OWC channel models based on deterministic and stochastic methods, fail to address mobility induced ambient light, optical turbulence and road reflection effects on channel characterization. 
Therefore, alternative machine learning (ML) based schemes, considering ambient light, optical turbulence, road reflection effects in addition to inter-vehicular distance and geometry, are proposed to obtain accurate VVLC channel loss and channel frequency response (CFR). This work demonstrates synthesis of ML based VVLC channel model frameworks through multi layer perceptron feed-forward neural network (MLP), radial basis function neural network (RBF-NN) and Random Forest ensemble learning algorithms. Predictor and response variables, collected through practical road measurements, are employed to train and validate proposed models for various conditions. Additionally, the importance of different predictor variables on channel loss and CFR is assessed, normalized importance of features for measured VVLC channel is introduced. 
We show that RBF-NN, Random Forest and MLP based models yield more accurate channel loss estimations with 3.53 dB, 3.81 dB, 3.95 dB root mean square error (RMSE), respectively, when compared to fitting curve based VVLC channel model with 7 dB RMSE. Moreover, RBF-NN and MLP models are demonstrated to predict VVLC CFR with respect to distance, ambient light and receiver inclination angle predictor variables with 3.78 dB and 3.60 dB RMSE respectively.

\end{abstract}


\IEEEpeerreviewmaketitle

\section{Introduction}

\Ac{VVLC} is a promising communication technology, aiming simultaneous data transmission and illumination through vehicle \ac{LED} lights.
\ac{VVLC} is considered as a secure complementary technology to \ac{RF} communications due to its \Ac{RF} interference free nature, license free wide spectrum availability, non-frequency selective flat fading, and directional \ac{LoS} channel characteristics.

Fundamentally, wireless communication channel models can be classified into two categories : deterministic and stochastic channel models. 
A deterministic channel model aims to predict the channel characteristics in a specific location with respect to transmitter and receiver locations, as well as the surrounding environment, exploiting computational electromagnetics with ray tracing and finite-difference time-domain (FDTD) methods. However, deterministic channel models lack computational efficiency and heavily depend on site-specific geometry with dielectric properties of scatter materials. 
On the other hand, stochastic approaches are utilized to reproduce the statistical behaviors of the channel yielding non site-specific but lower accuracy models. Stochastic channel models are classified into \ac{GBSM} where ensemble of the scatterers are placed in different geometrical positions based on statistical distributions and non$-$\ac{GBSM} fit measured or generated channel parameters into certain probability distributions. 
Stochastic approaches confront challenges with incorporating all relevant channel features. \ac{GBSM} highly depend on the probabilistic distributions of physical parameters (i.e. transmitter - receiver distance, angle, scatterer locations), whereas non-\ac{GBSM} stochastic channel models lack instantaneous channel spatial consistency, as all channel parameters in one channel realization are generated for a single location \cite{raschkowski2015metis}. Considering the required effort to obtain various probabilistic distributions for different environments, stochastic channel models generally rely on certain assumptions, where channel parameters are fitted, averaged out for generalized environments and possible scenarios, lacking precision. Moreover, mathematical expressions and probability distribution fitting constraints impose additional assumptions on the stochastic models, leading limited accuracy. 

Recently, \ac{ML} methods are proposed for channel modelling to overcome site-specific, high complexity limitations of deterministic approaches and low accuracy limitations of stochastic models \cite{aldossari2019machine}. Moreover, highly complicated mediums such as in-body, underwater, \ac{V2V} \cite{ramya2019using}, optical and molecular communication channels \cite{lee2017machine}, inherent certain distortion effects which are challenging to be expressed analytically. Therefore, \ac{ML} based channel modelling aims to develop low-complexity and accurate models for complicated channels, through direct learning of the robust patterns in the data without imposing any assumptions on the analytical expressions. Moreover, \ac{ML} models distinguish over scenarios by providing physical parameters corresponding to the specific scenario as inputs. 

Channel modelling with \ac{ML} can be classified into supervised and unsupervised learning based with respect to the labels of training data. Supervised learning based channel modelling, aims to learn a general function between inputs and outputs yielding solution for regression problems such as path loss predictions through labeled data\cite{zhang2019path,wen2019path}. On the other hand, unsupervised learning based channel modelling is favorable for clustering of \glspl{MPC} with same features such as delay, angle of arrival, and angle of departure, where unlabeled large amount of data is utilized \cite{he2018clustering}.

To date, deterministic and stochastic \ac{VLC} channel characterization is investigated for indoor\cite{al2016characterization,lee2011indoor, chvojka2015channel}, underground mine \cite{wang2018general,wang2017path} and outdoor \cite{lee2012evaluation,elamassie2018effect,eldeeb2019path,luo2015performance,cui2012traffic,al20182,chen2016time,cheng2017comparison,turan2018vehicular} environments mainly through simulation based studies. 

Deterministic \ac{VLC} channel models are investigated through ray tracing \cite{lee2012evaluation,elamassie2018effect,eldeeb2019path}, recursive methods \cite{luo2015performance,wang2018general,wang2017path,schulze2016frequency} and empirically \cite{cui2012traffic,chen2016time,cheng2017comparison,turan2018vehicular,memedi2017impact} with site-specific measurements. Ray tracing based \ac{VLC} channel modelling, yields \ac{CIR} with respect to the detected power of each ray and path lengths from source to detector considering reflections. 
To date, ray tracing has been used to extract \ac{VLC} channel delay profiles for \ac{ITS} applications in \cite{lee2012evaluation}, \ac{CIR} for \ac{VVLC} channel under fog and rain conditions in  \cite{elamassie2018effect}, path loss model for \ac{VVLC} in \cite{eldeeb2019path}. On the other hand, recursive methods yield \ac{PDP} where \ac{LoS} response is computed first and multiple bounces of light through reflecting elements are computed recursively with the assumption of same reflective characteristics (i.e. lambert surface) for all reflectors and scatterers. In the \ac{VLC} recursive channel modeling literature, indoor \ac{VLC} received power and time dispersion parameters are obtained through time domain simulations in \cite{lee2011indoor} and frequency domain simulations in \cite{schulze2016frequency}. Moreover, in the recursive channel model literature, \ac{VVLC} channel received power is modeled under the consideration of realistic headlight pattern and road reflection conditions in \cite{luo2015performance}, underground mine \ac{VLC} \ac{LoS} channel path loss and shadowing parameters are extracted with respect to Mie scattering and diffraction in \cite{wang2018general}, and \ac{NLoS} channel path loss model based on lambertian radiation pattern is proposed in \cite{wang2017path}. Empirical \ac{VLC} channel models further investigate real world effects on \ac{VLC} channel through measurements. Empirically, authors in \cite{cui2012traffic} provided a path loss model for traffic light to vehicle \ac{VLC} link including background solar radiation effects for a sunny day at specific location. Furthermore, \cite{chen2016time,cheng2017comparison} provided VVLC channel coherence time, auto correlation function and received power with respect to vehicle movements on a 18 km pre-defined route, driven 5 times. \cite{turan2018vehicular} provided VVLC channel time dispersion parameters through frequency domain channel sounding for specific time of day and location. Memedi \textit{et. al.} derived an empirical VVLC channel path loss model through stationary night measurements for unique environment, yielding \ac{RSS} with respect to transmitter $-$ receiver angle and distance \cite{memedi2017impact}. However, considering core dependency of \ac{VVLC} channel model on channel loss due to ambient light and atmospheric effects (i.e. fading, scintillation) with mobility, ray tracing based studies lack consideration of solar radiation and optical turbulence effects on \ac{VLC} link. Moreover, recursive methods rely on assumptions such as lambertian reflection for all surfaces while empirical models represent only the small portion of usage scenarios for deterministic models.

On the other hand, stochastic channel models offer increased flexibility, reduced computational complexity, and lower accuracy when compared to deterministic approach. \ac{VLC} and \ac{OWC} stochastic channel models are explored in \cite{perez1997statistical,al20182,chvojka2015channel}. Considering Non-\ac{GBSM}, \cite{chvojka2015channel} investigated impacts of mobility on indoor \ac{VLC} channel with respect to probabilistic movements for shadowing and blocking, yielding \ac{CDF} of \ac{RSS} for a range of people density. Moreover, authors in \cite{perez1997statistical} proposed a statistical method to obtain \ac{CIR} where Rayleigh and Gamma distributions are utilized to fit measured \ac{CIR} for an indoor \ac{OWC} channel. Under the consideration of \ac{GBSM}, \cite{al20182} proposed a 2D Non-Stationary \ac{GBSM} to generate \ac{CIR} for VVLC channel obtained Gaussian distributions for channel gain and \ac{RMS} delay spread.

Considering the limited applicability of deterministic and stochastic methods on \ac{VVLC} channel models due to site-specific characteristics and low accuracy, \ac{ML} based channel model frameworks with the capability of learning complex features pave the way to accurately model \ac{VVLC} channel propagation.  

To date, \ac{ML} based \ac{OWC} channel modelling has not been investigated in the literature. However, \ac{ML} based channel model frameworks trained through measurement data sets, on contrary to relying numerous assumptions are proposed for \ac{RF} communications \cite{ramya2019using,huang2018big,bai2018predicting,ostlin2010macrocell,ferreira2016improvement,ayadi2017uhf,zhang2019real,lee2017machine}. Considering \ac{mmWave} communication channels, Huang \textit{et.al.} proposed an \ac{ANN} enabled channel model framework, to obtain channel parameters including received power, \ac{RMS} delay and angle spreads achieving 1.64 dB \ac{RMSE} for received power estimations \cite{huang2018big}. Authors in \cite{bai2018predicting} proposed \ac{CNN} based three-dimensional \ac{mmWave} massive \ac{MIMO} channels framework yielding 0.34 dB to 3.05 dB \ac{RMSE} for channel path loss with transmitter and receiver location inputs under the consideration of 5 different ray tracing based data sets. For mobile channel modelling, path loss predictions through \ac{ANN} is demonstrated to outperform statistical and Okumura–Hata model with maximum error of 22 dB, mean error of 0 dB with 7 dB standard deviation in \cite{ostlin2010macrocell}. \ac{ANN} aided hybrid signal strength prediction at 1140 MHz is depicted to provide 8 dB average improvement when compared to pure ITU-R.526-11 model in \cite{ferreira2016improvement} whereas, an \ac{ANN} based propagation model for 450, 850, 1800, 2100, and 2600 MHz yields path loss predictions with 0.235 dB absolute mean error in \cite{ayadi2017uhf}. For vehicular communication channels, authors in \cite{zhang2019real} proposed an \ac{ANN} based channel model outperforming generalized gamma, polynomial fitting and dual slope distance-break point models for path loss predictions. Ramya \textit{et.al} showed that non-parametric learning based Random Forest method, increased \ac{V2V} channel path loss prediction accuracy with 2.2 dB mean and 1.5 dB standard deviation of absolute error when compared to log distance path loss model \cite{ramya2019using}. On the other hand, \cite{lee2017machine} proposed an \ac{ANN} based molecular communications channel model to predict channel model parameters accurately, where obtaining exact analytical channel model is challenging. Thereby, \ac{ML} can be regarded as an appealing approach for accurate and computationally efficient wireless communication channel modelling.

Taking into account the considerable amount of work in the \ac{VLC} channel modelling literature, none of the studies to date, characterized \ac{VVLC} links targeting \ac{V2V} with respect to various ambient light, exhaust plume induced optical turbulence, inter-vehicular distance, receiver inclination angle, lane occupancy conditions and \ac{LED} frequency response through practical road measurements. Hence, this work presents \ac{ML} based frameworks to extract \ac{VVLC} channel signal attenuation as a function of distance, \ac{LoS} - \ac{DLoS} conditions through receiver inclination angle, \ac{LED} modulation frequency, occupied lane, optical turbulence and ambient light. \ac{VVLC} channel frequency response and path loss measurements conducted via production vehicle \ac{LED} lights in real road scenarios are utilized to train and validate \ac{ML} based models. Proposed channel model frameworks are directly learned from measurement data sets yielding higher accuracy than slope intercept fits proposed for \ac{VLC} channel. Furthermore, a comparative study between Random Forest non-parametric learning method and two types of \glspl{NN}, \ac{MLP}, \ac{RBF} is conducted to model \ac{VVLC} channel path loss. The goal is to obtain a \ac{ML} based \ac{VVLC} channel propagation model framework that is not overly complex but still generalizes well and is accurate enough for practical \ac{VVLC} applications. In this work we leverage the use of \ac{ML} techniques towards channel path loss and channel frequency response estimation, to accurately model \ac{VVLC} channel with respect to given physical conditions. In particular, we propose exploiting \ac{ML} to predict \ac{VVLC} link quality depending on features of ambient light, inter-vehicular distance, transmitter receiver geometry, and modulation frequency, to enable better utilization of the \ac{VVLC} channel.



\captionsetup[figure]{font=footnotesize,labelfont=footnotesize} 
\begin{figure}[h]
\centering
\includegraphics[width=\linewidth]{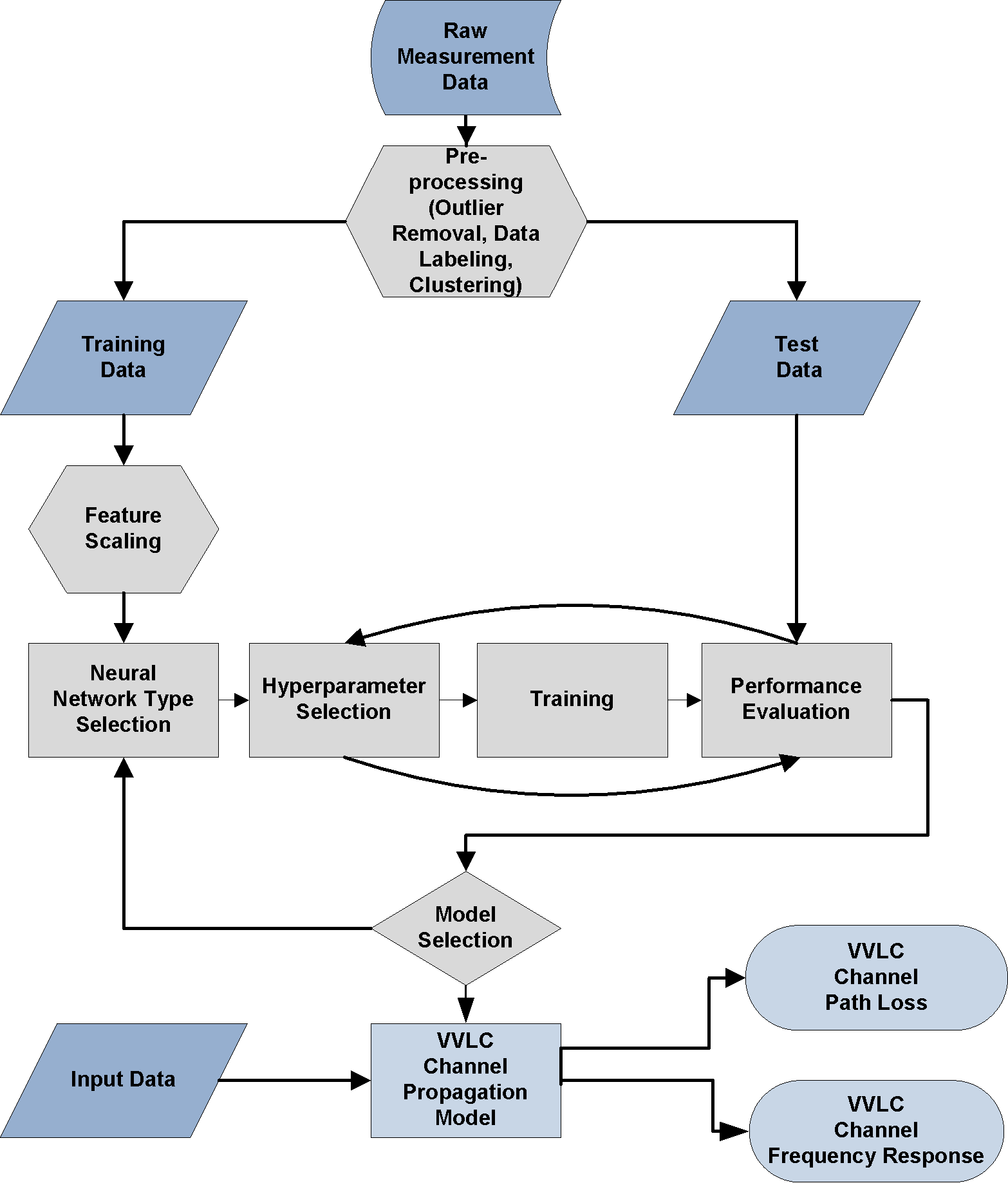}
\caption{ML Based VVLC Channel Propagation Modelling.}
\end{figure}

The main novelties and contributions of this work are as follows. 
\begin{itemize}
  \item We extracted the importance of predictor variables to obtain \ac{VVLC} channel loss and \ac{CFR} through measured \ac{VVLC} propagation channels. We fit the measured channel loss data to current \ac{VLC} channel models, and demonstrated that the current models do not capture channel loss deviations due to mobility and environment induced variations. This is the first work to provide a measurement based quantitative analysis of \ac{VVLC} channel under various ambient lighting, optical turbulence through exhaust plume, inter-vehicular distance, and geometry with both \ac{LoS} and \ac{DLoS} conditions. 
  \item We proposed \ac{MLP}, \ac{RBF} \ac{NN} and Random Forest learning based channel model frameworks to be trained with the predictor variables of inter-vehicular distance, geometry, ambient light, optical turbulence existence, lane occupation and receiver inclination angle features yielding highly accurate channel loss and channel frequency response for the cases that proposed frameworks were not trained with. This is the first study to propose \ac{ML} based \ac{VVLC} channel propagation models.   
  \item We evaluated the validity and performance of the proposed models with their sensitivity to the amount of training data. This is the first work to analyze the validity and robustness of \ac{ML} based channel model frameworks for \ac{VVLC} channels across a wide range of varying physical conditions.
\end{itemize}

The remainder of this paper is organized in the following way. Section~\ref{challenges} outlines the \ac{VVLC} channel differences from \ac{RF} vehicular communication channels and other \ac{VLC} channels, highlighting the unique challenges of \ac{VVLC} channel modelling. Data collection details for \ac{VVLC} channel characterization and the features utilized to annotate measurement data are detailed in Section~\ref{CHMEAS}. Existing channel path loss models and their comparison to measurement data is provided in Section~\ref{EXISTING}. Section~\ref{Methods} introduced \ac{ML} based channel characterization methodology, where detailed system model is can be found in Section~\ref{sysmodel}. Performance evaluation and comparisons of the proposed channel model frameworks can be found in Section~\ref{Performance}. Finally, Section~\ref{conclusion} concludes the paper.

\section{Vehicular VLC Channel Modelling}
\label{challenges}

\subsection{Vehicular VLC Channel}
Vehicles are operated in varying weather, climate, illumination and road conditions. Mainly, atmospheric interaction yielding a combination of absorption and scattering plays an important role on \ac{VVLC} channel characteristics. \ac{VVLC} channel characterization and utilization further depends on optical turbulence, ambient light induced noise, transmitter-receiver geometry, and low pass frequency response characteristics of \glspl{LED} and optical receivers. 

Optical turbulence, sourced by random temperature fluctuations on road surface and around exhaust plumes distort \ac{VVLC} signals, resulting optical power fluctuations \cite{li2015effects}. However, accurate characterization of optical turbulence and finding a mathematical expression to incorporate into \ac{VVLC} channel model is challenging, due to mobility induced abrupt temperature, wind, weather changes. 

\captionsetup[figure]{font=footnotesize,labelfont=footnotesize} 
\begin{figure} [h]
    \centering
  \subfloat[\label{solar}]{%
       \includegraphics[width=0.5\linewidth]{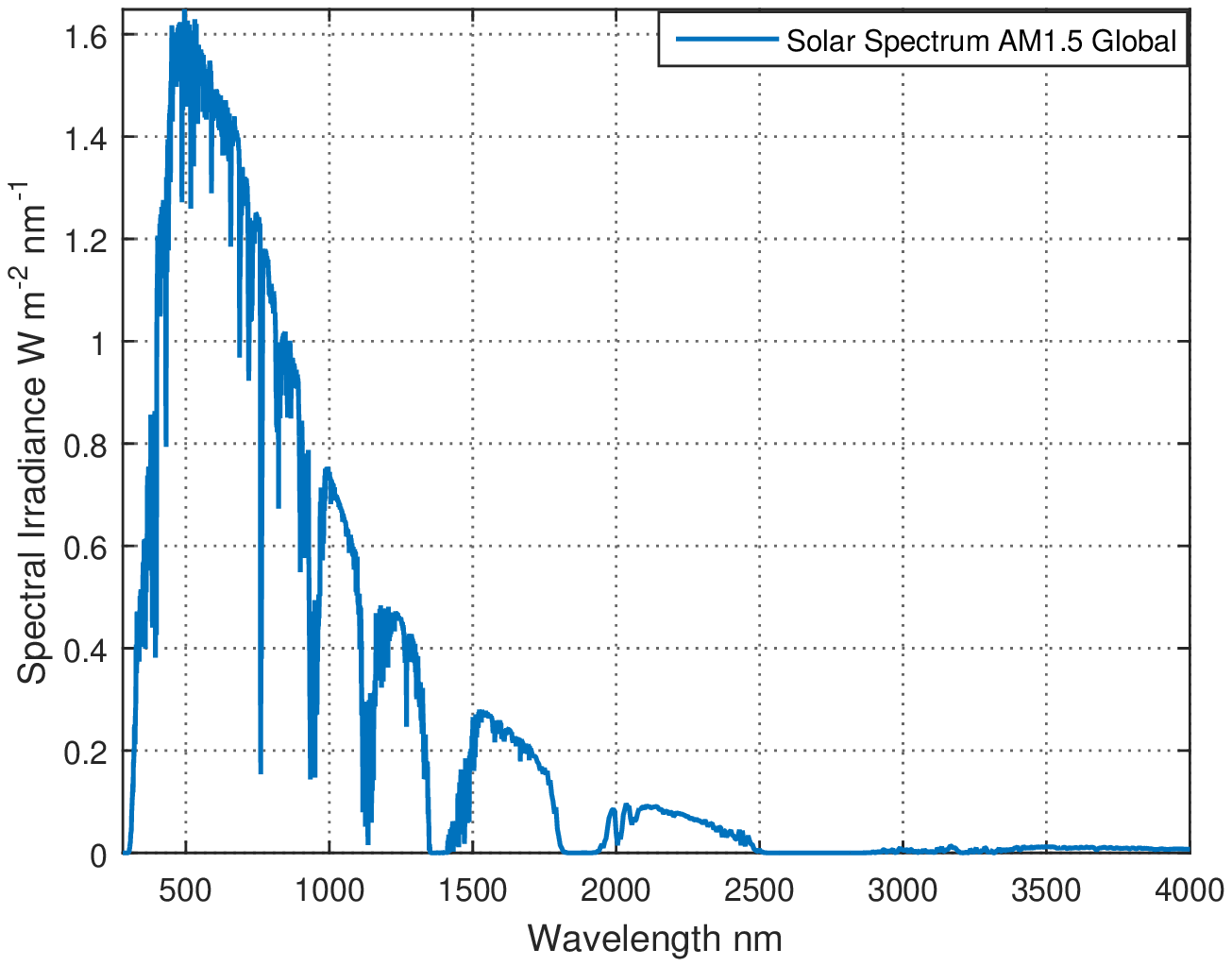}}
    \hfill
  \subfloat[\label{pdcomp}]{%
        \includegraphics[width=0.5\linewidth]{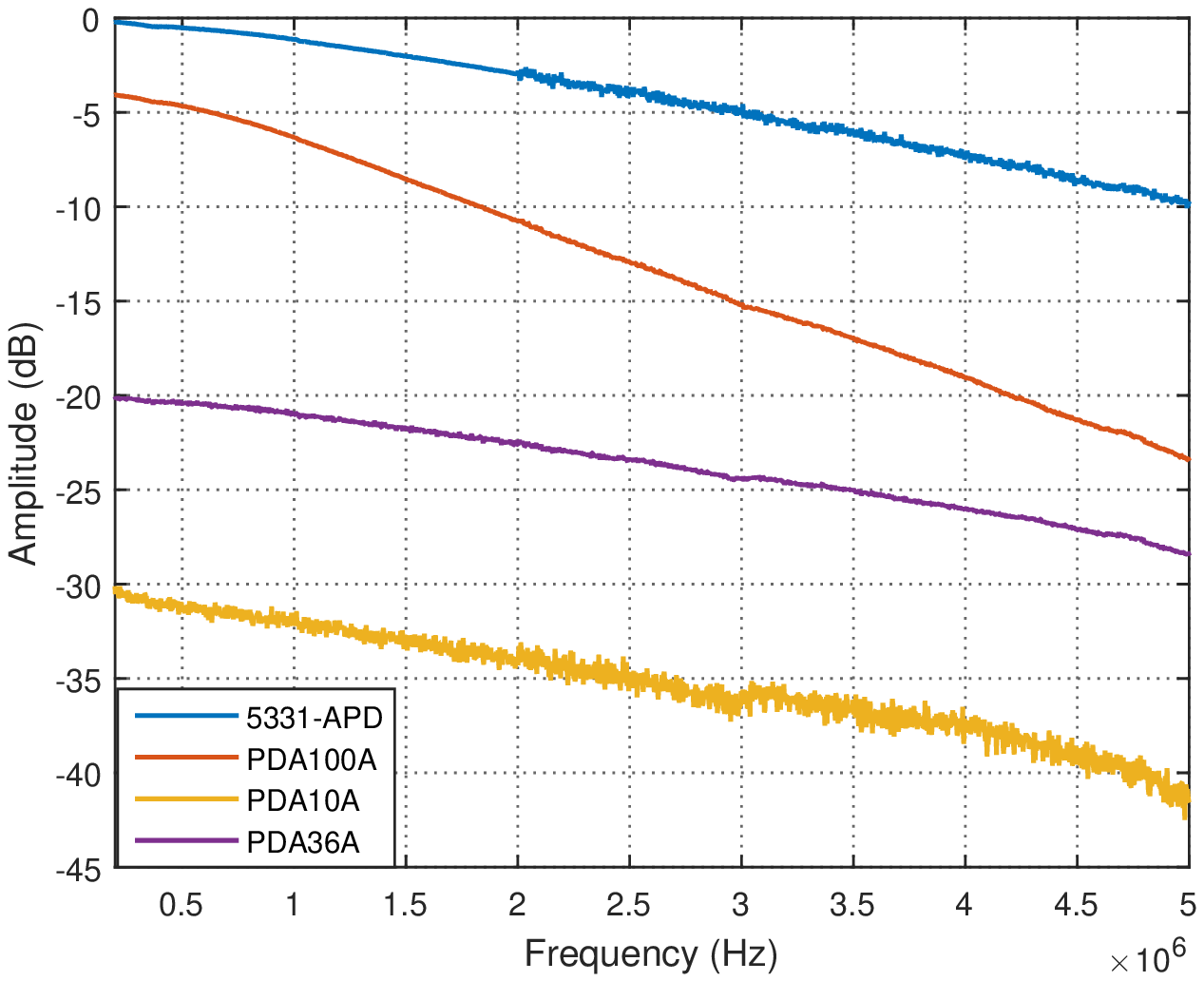}}
    \\
  \caption{(a) Solar Spectrum (b) Various Optical Detectors Frequency Response for Vehicle LED Headlight}
  \label{detectors}
\end{figure}

More photons from sun light reaches to optical detector during day time. Thus, the number of photons to reach the optical receiver from \ac{VVLC} transmitter decreases. Moreover, photons absorbed from ambient lights and solar radiation excite electrons and cause them to generate current in the form receiver shot noise and thermal noise. Sunlight contamination dominates the noise and determines the number of photons captured by the receiver for daylight conditions. As solar spectrum (See Fig.\ref{solar}) is stronger in the visible light region, optical sun interference filters, that attenuate communication signals are not favorable for practical \ac{VVLC} systems. Therefore, with the increase in ambient light, less photons from \ac{VVLC} signals reach to the receiver, and dynamic range of the \ac{VVLC} receiver decreases due to increased receiver noise, leading more channel loss. 

Fig.\ref{pdcomp} depicts the difference between various optical receivers (Hamamatsu C-5331 APD, Thorlabs PDA100A, Thorlabs PDA36A, Thorlabs PDA10A) with respect to same optical signal swept from 100 kHz to 5 MHz at fixed distance and ambient light conditions. It is clear that, due to receiver aperture, spectral responsivity and inherent gain stage of the receivers, \ac{VVLC} channel loss and frequency response depends on the receiver selection. Hence, \ac{VVLC} channel loss, incorporating receiver noise in addition to propagation loss due to atmospheric attenuation and fading should be considered for \ac{VVLC} channels.

\captionsetup[figure]{font=footnotesize,labelfont=footnotesize} 
\begin{figure}[h]
    \centering
  \subfloat[\label{ledtemp}]{%
       \includegraphics[width=0.5\linewidth]{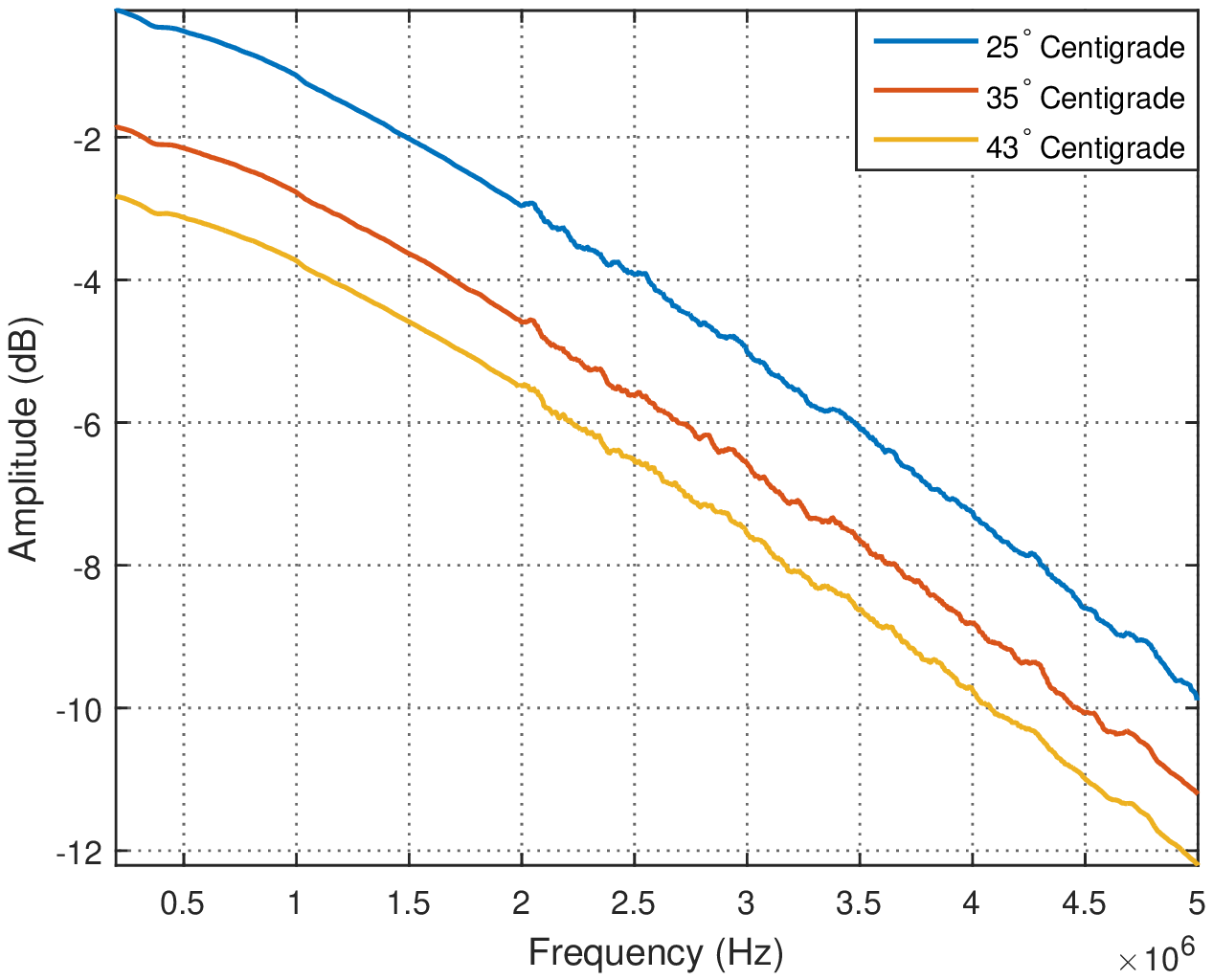}}
    \hfill
  \subfloat[\label{ledspectrum}]{%
        \includegraphics[width=0.5\linewidth]{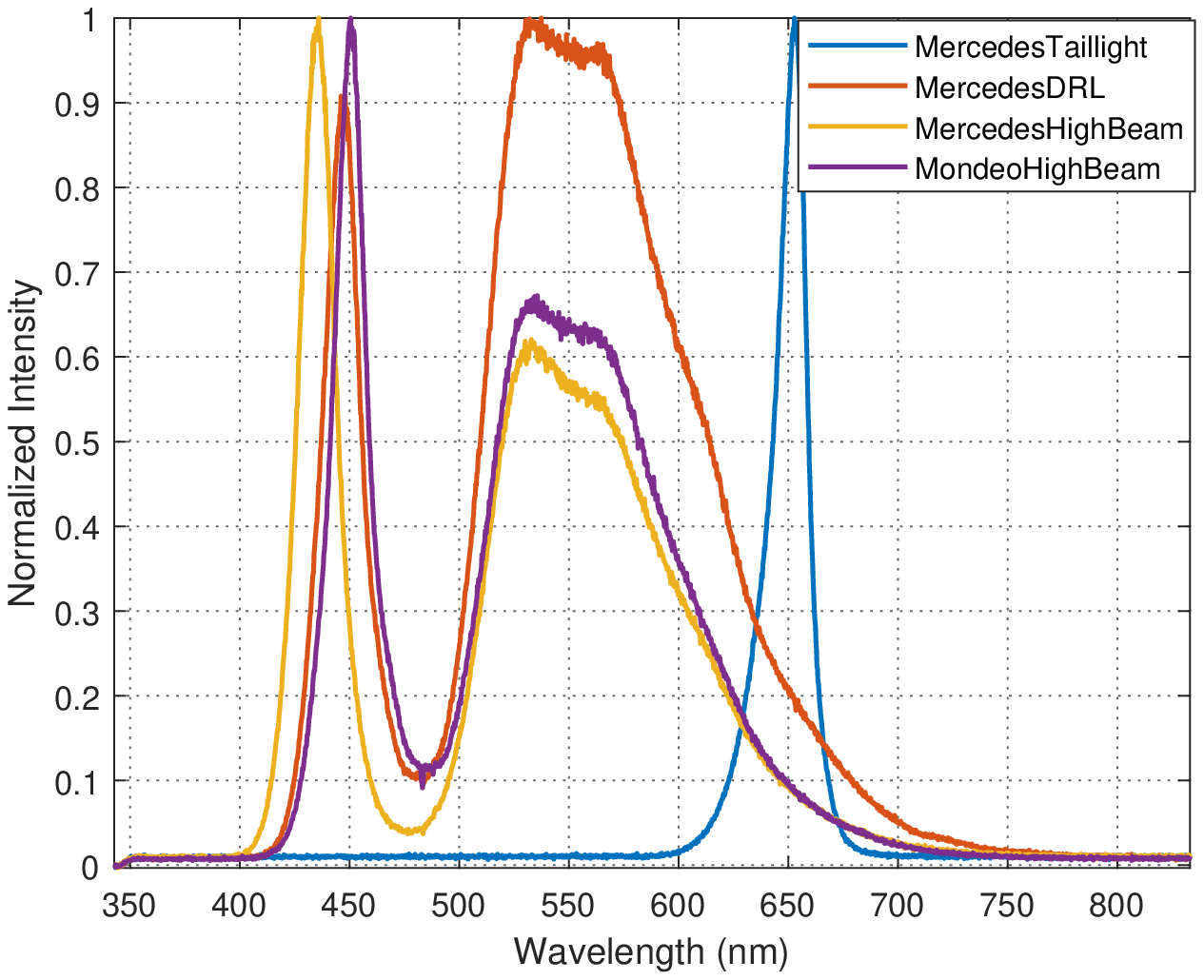}}
    \\
  \caption{(a) Production vehicle LED headlight frequency response degrades with the increasing temperature (b) Spectrum of various vehicle LEDs depicts the difference between LEDs of same purpose (i.e. headlight)}
  \label{LEDs}
\end{figure}

Vehicle \ac{LED} light half intensity beam angle (HIBA), regulation \cite{no2004112} driven minimum illumination \ac{FoV}, with the utilization of beam shaping optics such as lenses, reflectors and mirrors determine \ac{LoS} and \ac{DLoS} characteristics of \ac{VVLC} channel. Compared to indoor and underwater \ac{VLC} channels, diffuse components sourced from nearby scatters is weak for \ac{VVLC} channels. Hence, \ac{VVLC} channels are mainly explored for their \ac{LoS} characteristics. Low pass frequency response of both \glspl{LED} and optical receivers impose additional limitations on the \ac{VVLC} received optical signal power, as high frequency modulated signals are transmitted and received with lower optical power. \ac{LED} temperature dependent characteristics \cite{colaco2017thermal} (See Fig.\ref{ledtemp}) and varying spectral properties of each automotive \ac{LED} (See Fig.\ref{ledspectrum}) further determines VVLC link performance and channel loss.

Multipath fading poses limitation on \ac{RF} based vehicular communication system performance due to different transmitter and receiver geometries, mobility, nearby scatter objects, and dynamic propagation conditions. However, \ac{VVLC}, utilizing non-coherent vehicle \ac{LED} lights, and receiver apertures in the order of millions of wavelengths, is immune to multipath fading with no small fading \cite{al2018optical}. Even though, multipath fading is not a major concern for \ac{VVLC}, multipath dispersion, sourced from the transmitted signals outreach to the receiver via different paths and times cause symbols spreads, yielding \ac{ISI}. Road surface with different reflection properties and nearby scatters such as guard rails and vehicles can be considered as main source of multipath dispersion for \ac{VVLC}. However, considering the low amplitude of \glspl{MPC} when compared to strong \ac{LoS} signals, multipath dispersion has subtle \ac{ISI} effects on practical \ac{VVLC} inter-vehicular distances (i.e. 10 m to 100 m). 

On the other hand, high Doppler spread in vehicular environments for \ac{RF} based vehicular schemes, induces short channel coherence time, requiring accurate channel estimation for reliable communications. However for \ac{VVLC}, considering $650nm$ wavelength taillight \ac{LED}, at a vehicle speed of 250km/h, yields  $210MHz$ Doppler frequency or $0.00015nm$ wavelength shift from its nominal value. Since the optical receiver detects only the intensity of the optical wave, the generated photo current will deviate from the expected level based only on the spectral sensitivity (A/W) of the receiver. Even though this results with electrical \ac{SNR} variation at the receiver side, as the wavelength shift due to Doppler spread is subtle, Doppler spread can be regarded negligible for \ac{VVLC} channels. 

\ac{VVLC} channels exhibit unique characteristics among \ac{V2X} communication channels, in terms of their ambient light and atmospheric dependencies where analytical characterization and generalizations lead challenges. 

\begin{figure}[h]
\centering
\includegraphics[clip,width=\linewidth]{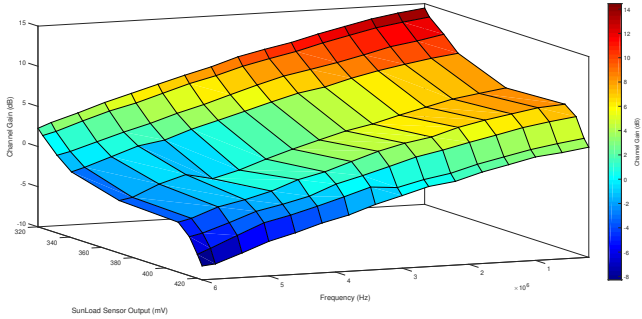}
\caption{Channel Gain at $6$m under Varying Ambient Light}
\label{CGF}
\end{figure}

\begin{figure*}
\centering
\includegraphics[width=\textwidth]{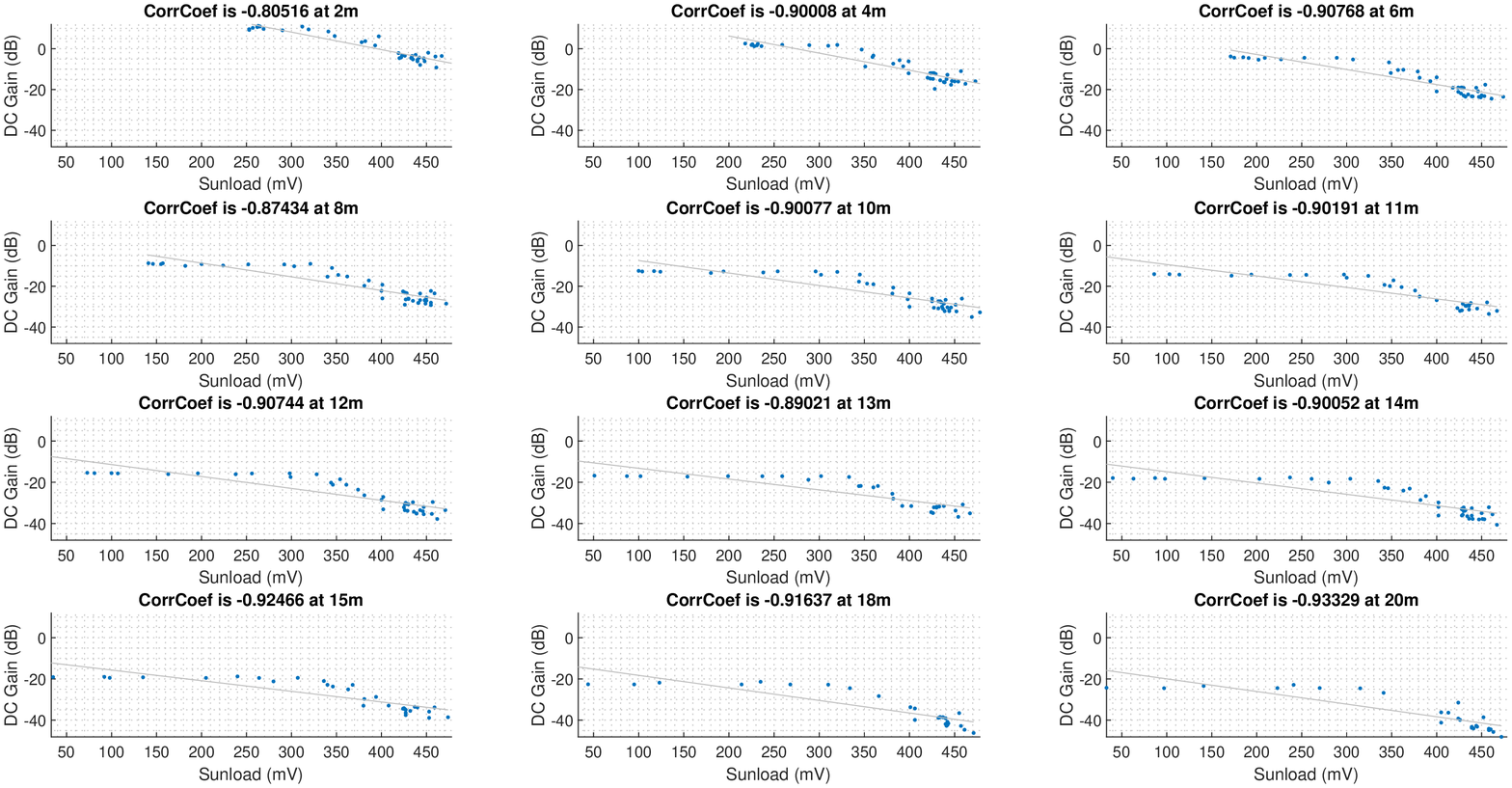}
\caption{DS1 Ambient Light and Channel Gain Correlation at Various Distances}
\label{ambient}
\end{figure*}

Fig. \ref{CGF} shows the relation between \ac{LED} modulation frequency, ambient light and channel DC gain at the inter-vehicular distance of 6 m, where channel DC gain is observed decrease with increasing ambient light and modulation frequency. Fig.\ref{ambient} represents the correlation between channel loss and ambient light for each measurement distance of \ac{LoS} \ac{VVLC} channel. Unlike \ac{RF} \ac{V2X} channel models, characterization of  Doppler Spread and multipath fading is not targeted for \ac{VVLC} channel characterizations. Moreover, ensuring directional long distance optical communications, channel loss characterization poses more importance than time dispersion parameter extraction, which is not the case for indoor \ac{VLC} channels.

\captionsetup[figure]{font=footnotesize,labelfont=footnotesize}

\section{Channel Measurements and Measured Data Statistics}
\label{CHMEAS}
\subsection{Channel Loss Measurements}

Two different measurement campaigns are executed to capture \ac{VVLC} channel properties. Static frequency domain \ac{VVLC} channel sounding is conducted to obtain channel frequency response with respect to varying inter-vehicular distance and ambient light conditions. On the other hand dynamic \ac{RSS} based measurement campaigns targeted characterization of \ac{VVLC} channel loss for various ambient light, receiver angle, optical turbulence region and lane occupation conditions. Moreover, dynamic \ac{RSS} based channel loss measurements aim to investigate the effects of both \ac{LoS} and \ac{DLoS} propagation where receiver is inclined towards road surface to capture optical signals both from \ac{LoS} and road surface reflections. Two different data sets of \ac{DS1} for \ac{CFR} and \ac{DS2} for channel loss characterizations. Data sets \ac{DS1} and \ac{DS2} are formed through field measurements of 29631 and 61488 samples respectively.

\subsubsection{Frequency Domain Measurements}

Frequency domain \ac{VVLC} channel measurements are handled by a closed loop \ac{VNA} based approach to accurately characterize \ac{VVLC} channel path loss dependence on modulation frequency. At the transmitter, Port 1 of Rohde \& Schwarz ZNB20B or Anritsu 2026-C \ac{VNA} is connected to a $47dB$ amplifier which consists of two cascaded \glspl{LNA} considering their $1dB$ compression point. The output of the amplifier is connected to bias-tee where the DC-bias voltage is selected at the linear working region of the \ac{LED}. The resulting signal is fed to a MY2017 Ford Mondeo \ac{LED} headlight. At the receiver, Hamamatsu S3884 $-$ C5331 \ac{APD} is utilized to capture the optical signal. The output of the utilized photo detector is connected to a 25dB Mini Circuits ZFL500 \ac{LNA} to increase \ac{SNR} of the captured signal. Amplifier output is fed to Port 2 of the \ac{VNA} using three Huber-Suhner Sucoflex 404 shielded microwave cables. 

The measurements are taken under the same calibration with same cables and connectors. The \ac{VNA} is operated at $-20dbM$ output power mode. For all measurements, $N_f$=4001 samples are recorded with 5 averaging for each sweep to reduce random noise. \ac{IF} bandwidth is selected as 500 Hz, enabling -70 dB noise floor level. 

{$S_{21}$} parameter measurements at 1411 different points are taken for \ac{LED} modulation frequencies between 2 kHz - 10 MHz from $2m$ (i.e. bumper to bumper traffic) to $20m$ (i.e. platoon distances) distances at various background light levels including outdoor sunny day, night time with ambient lights on and off, sunrise, sunset, and cloudy weather. However, considering limited modulation bandwidth of vehicle \ac{LED} light under interest, measurement points between 2 kHz - 2 MHz with 100 kHz step sizes are employed for channel frequency response and channel loss characterization forming \ac{DS1}.   

Path loss measurement campaign yielded, frequency dependent path loss ($-S_{21}$) of VVLC channel with respect to inter-vehicular distance and sun load sensor outputs. DS1 is composed of 29631 samples of 21 variables, where 1411 measurement points are considered with 19 channel loss magnitude variables for \ac{LED} modulation frequencies between 2 kHz- 2 MHz (100 kHz intervals), with distance and sun load sensor predictor variables.

\subsubsection{RSS Measurements}
For dynamic \ac{VVLC} channel \ac{RSS} measurements, Rohde Schwarz FSV-3 \ac{VSA} with Hamamatsu S3884 $-$ C5331 \ac{APD} is employed at the receiver vehicle where 1 MHz sinusoidal tone generated from \ac{AWG} is fed to a vehicle \ac{LED} \ac{DRL} through \ac{LED} current driver at the transmitter vehicle. Measurements up to $114$ m inter-vehicular distances are captured through remote control of \ac{VSA} with LabView software where \ac{GPS} locations are fed from \ac{GPS} disciplined oscillator of NI USRP 2932 software defined radio, accelerometer and production vehicle sun load sensor voltage values are recorded with NI MyRIO real time embedded controller. Considering limited accuracy of \ac{GPS} receiver, Velodyne VLP-16 Lidar is utilized for distance measurement and range validations. Our \ac{VVLC} channel sounding setup is detailed in \cite{turan2018vehicular}.

As \ac{VVLC} channel \ac{RSS} also depends on angular variations sourced through road surface (i.e. bumps), accelerometer values are recorded to observe road surface and driving style dependent \ac{VVLC} signal \ac{RSS} fluctuations where extreme variations are considered to be outliers. Dynamic measurements are composed of four different scenarios including \ac{LoS} same lane leader follower, \ac{DLoS} same lane leader follower, \ac{LoS} next lane leader follower, \ac{DLoS} next lane leader follower, where receiver inclination angle is 30~\degree for \ac{DLoS} conditions to better capture road surface reflections and decrease sun light interference, similar to production vehicle's rear view camera orientations. Either transmitter or receiver vehicle is located at a fixed location while the other is moved with a maximum velocity of 10 km/s up to 114 m distance during dynamic scenario measurements. For dynamic scenarios 7686 \ac{RSS} values are captured. Measurement setup specifications are summarized in Table~\ref{OFE}. \ac{DS2} is composed of 61488 measurement samples, where distance, ambient light, occupied lane and receiver angle are predictor variables, and channel loss is the response variable, x, y, z axis acceleration measurements are validation variables to detect outlier samples.

The features we annote \ac{RSS} and \ac{CFR} measurements are, 
\begin{enumerate}
  \item \textbf{Intervehicular Distance:} (a number in meters 50 cm - 114 m) the distance between the transmitter and receiver vehicles, calculated from GPS locations, laser distance finder and LiDAR point clouds.
  \item \textbf{Ambient Light:} (a number in millivolt) voltage values increases with the solar radiation and ambient light, where the value changes between 33 (complete darkness) to 475 (sun shine, clear sky). 
  \item \textbf{Receiver Inclination Angle:} (a number in degrees) the optical receiver elevation angle, varied between 0 \degree for \ac{LoS} and 30 \degree for \ac{DLoS}.
  \item \textbf{Occupied Lane:} The vehicles are either located in the same lane denoted as 1, or nearby lane denoted with 0.
  \item \textbf{Optical Turbulence:} Optical turbuence sourced by vehicle exhaust is observed to be substantial when receiver vehicle equipped with optical detector at the rear of the vehicle moves in reverse direction, due to exhaust plumes scattering, optical turbulence existence is labeled as 1 whereas the non-existence is denoted with 0.  
  \item \textbf{Variance Region:} Nearby distance amplitude measurements are observed to have high variance, where k-means clustering unsupervised \ac{ML} algorithm detailed in Section~\ref{sysmodel}-C is utilized to label high variance region with 1, and the rest as 0. 
  \item \textbf{VNA Model} Two different \glspl{VNA} are utilized for \ac{CFR} measurements, labeled as 1 or 0. $S_{21}$ parameter amplitude differences are observed due to varying \ac{LED} driving capability and non-accurate calibrations of \glspl{VNA}, regarding low impedance \ac{LED} loads.
  
\end{enumerate}

\begin{table}[]
\centering
\caption{Measurement Setup Front End Specifications}
\label{OFE}
\begin{tabular}{lc}
\hline 
\textbf{Parameter} & \textbf{Value} \\ \hline
\multicolumn{2}{c}{\textbf{Transmitter}} \\ \hline
Headlight 3-dB Bandwidth & 2 MHz \\ 
DC Bias Voltage & 24 $V$ \\
Driver Block Input Signal Amplitude & 63 m$V_{pp}$ \\
Driver Block Total Gain  & 47 dB  \\
Driver Block Output Signal Amplitude& 14.1 $V_{pp}$ \\
LED Input Signal Amplitude & 5.6 $ V_{pp}$ \\ 
LED Optical Transmitted Power & -6.72 dBm \\ 
Transmitter Height & 0.7 cm \\ \hline
\multicolumn{2}{c}{\textbf{Receiver}}\\ \hline
Avalanche Photodiode Module & Hamamatsu C5331-03\\
APD Active Area & 1 $mm$\\
APD 3 dB Frequency Bandwidth & 4kHz to 100 MHz \\ 
APD Spectral Response Range & 400 to 1000 nm \\
APD Peak Sensitivity Wavelength & 800 nm\\
Amplifier & Mini-Circuits ZFL-1000LN+\\
Amplifier Gain & 20 dB \\
Amplifier Frequency Range & 0.1 to 1000 MHz \\
Receiver Height & 0.7 m\\ \hline \hline
\end{tabular}
\end{table}





\begin{table}[]
\caption{Measurement Data Statistics}
\begin{adjustbox}{width=\linewidth}
\begin{tabular}{|l|l|l|l|l|l|l|c|c|}
\hline
\multicolumn{1}{|c|}{\multirow{3}{*}{\textbf{\begin{tabular}[c]{@{}c@{}}Measurement\\ Data\end{tabular}}}} & \multicolumn{6}{c|}{\textbf{Path Loss (dB)}}                                           & \multirow{3}{*}{\textbf{Max Distance (m)}} & \multirow{3}{*}{\textbf{\begin{tabular}[c]{@{}c@{}}Number of \\ Path Loss / RSS \\ Measurement Points\end{tabular}}} \\ \cline{2-7}
\multicolumn{1}{|c|}{}                                                                                     & \multicolumn{2}{c|}{\textit{200 kHz}} & \multicolumn{2}{c|}{\textit{1 MHz}} & \multicolumn{2}{c|}{\textit{2 MHz}} &                                            &                                                                                                                        \\ \cline{2-7}
\multicolumn{1}{|c|}{}                                                                                     & $\mu_{200k}$       & $\sigma_{200k}$       & $\mu_{1M}$         & $\sigma_{1M}$        & $\mu_{2M}$         & $\sigma_{2M}$        &                                            &                                                                                                                        \\ \hline
\begin{tabular}[c]{@{}l@{}}DS1- Night\\ (SL \textless 300 mV)\end{tabular}                                 & 7.82          & 13.07        & 8.98         & 13.27       & 11.09        & 13.66       & 20                                         & 998                                                                                                                    \\ \hline
\begin{tabular}[c]{@{}l@{}}DS1- Day\\ (SL \textgreater 300 mV)\end{tabular}                                & 20.25         & 13.84        & 21.59        & 14.11       & 23.83        & 14.21       & 20                                         & 413                                                                                                                    \\ \hline
DS2 (Day-Night)                                                                                            & \multicolumn{2}{c|}{-}       & 50.81        & 17.89       & \multicolumn{2}{c|}{-}     & 114                                        & 7686                                                                                                                   \\ \hline
\end{tabular}
\end{adjustbox}
\end{table}

\captionsetup[figure]{font=footnotesize,labelfont=footnotesize} 
\begin{figure*}[h]
\centering
\begin{minipage}[c]{.47\textwidth}
\centering
\includegraphics[trim={15 0 30 0},clip,width=\linewidth]{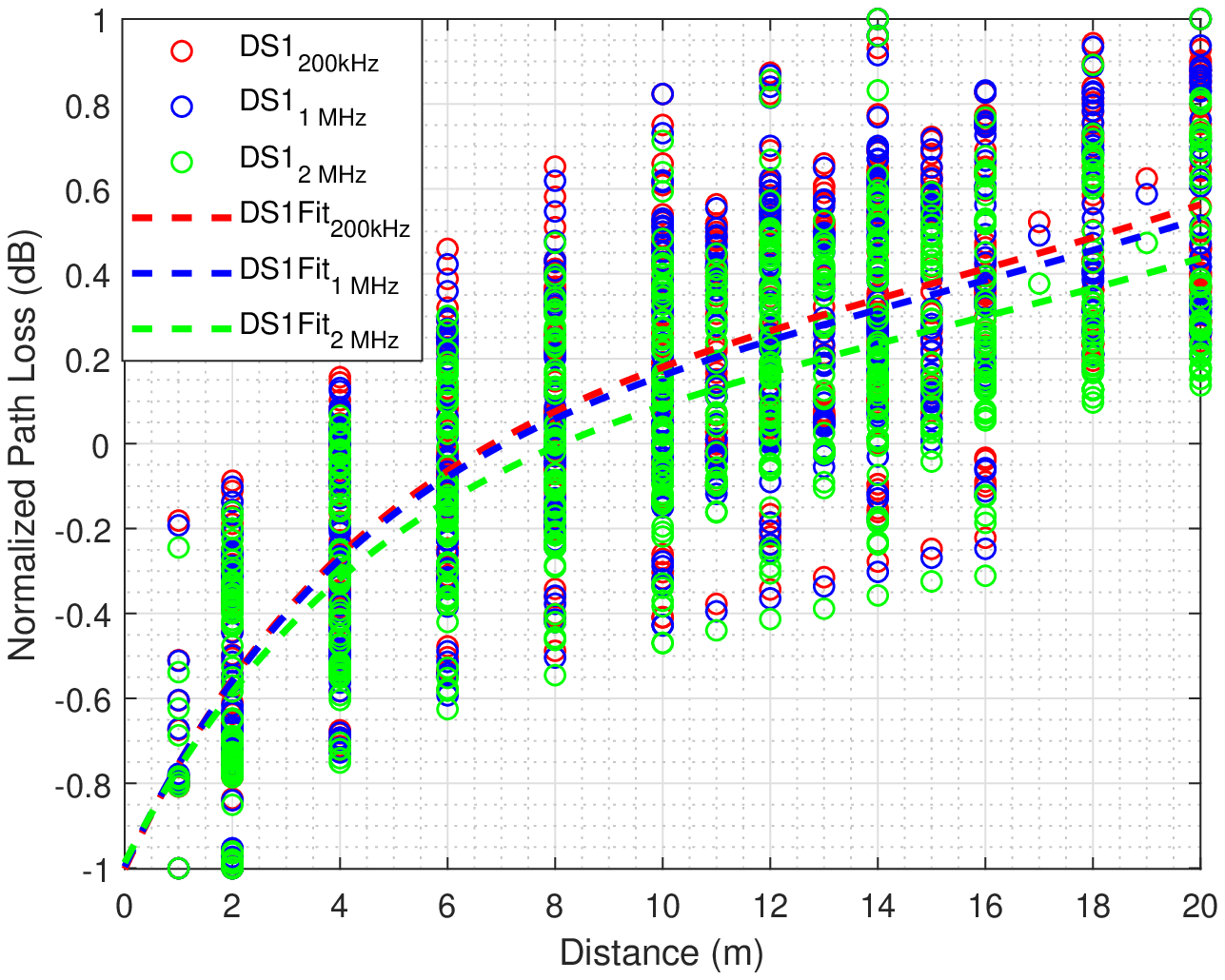}
\caption*{(a)}
\end{minipage}
\begin{minipage}[c]{.47\textwidth}
\centering
\includegraphics[trim={15 0 30 0},clip,width=\linewidth]{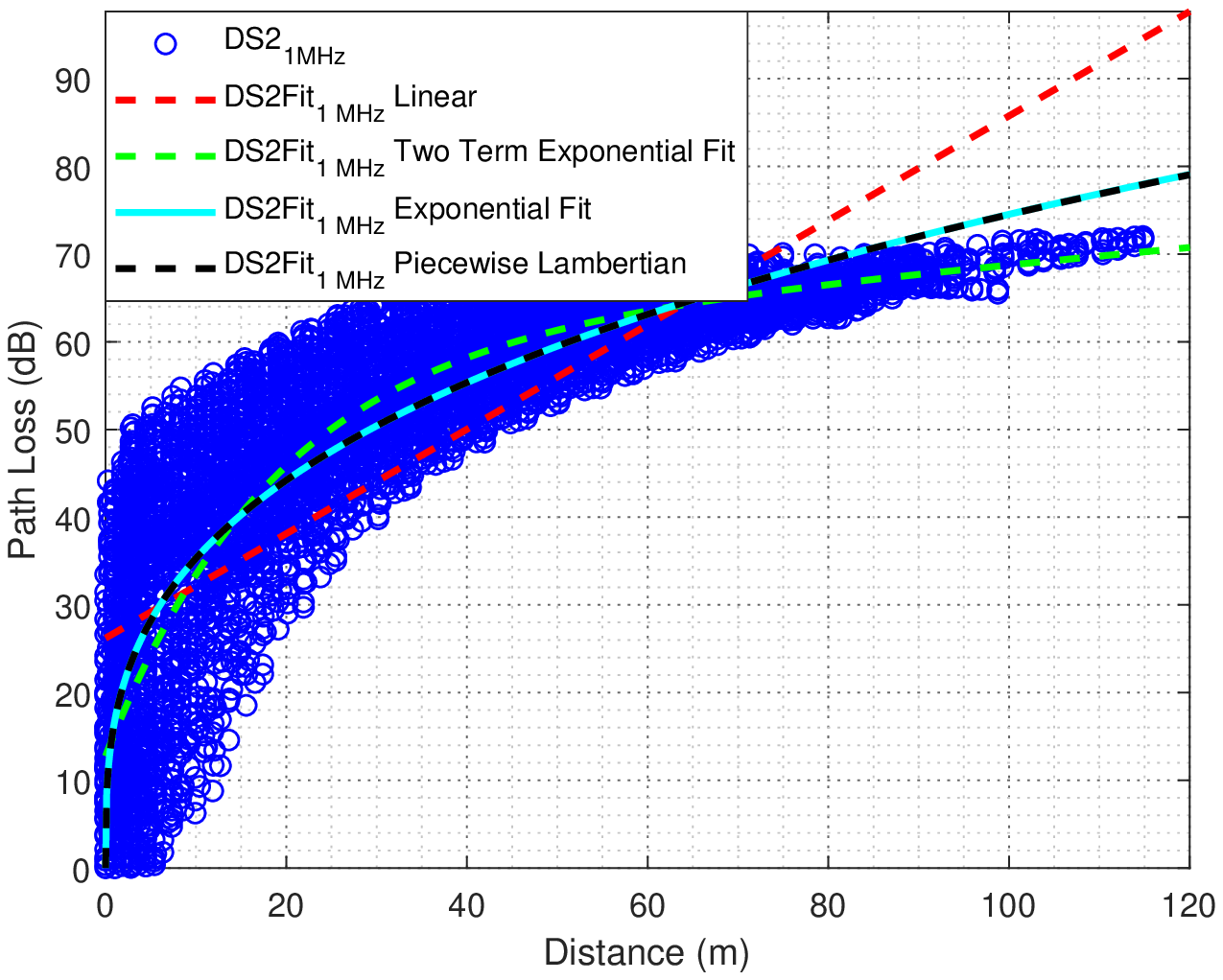}
\caption*{(b)}
\end{minipage}
\caption{(a) Dependence of Dataset 1 path loss on the distance and modulation frequency with two term exponential fit (b) Dependence of Dataset 2 path loss on the distance with linear, two term exponential, exponential and piecewise Lambertian fit}
\label{all_data}
\end{figure*}

\section{Existing Channel Path Loss Models and Comparison to Measurement Data}
\label{EXISTING}

The path loss between a \ac{VVLC} transmitter and receiver originates from free-space attenuation, scattering, angular orientation and ambient light induced receiver noise.

To date, \ac{VVLC} channel path loss is modeled with Lambertian model, linear fitting, exponential and two term exponential fitting. Lambertian model is modified with respect to vehicle optics at Piecewise Lambertian model \cite{viriyasitavat2013short} and channel DC gain is expressed as,

\begin{equation}
H(0)= \frac{(n+1)A}{2\pi D^{\gamma}}cos^{n}\phi  cos(\theta)
\end{equation}{}

where n is the Lambertian model order given as $n=-\frac{ln2}{lncos(\phi_{1/2})}$ , $\phi_{1/2}$ is the \ac{LED} half power angle, $\gamma$ is the path loss exponent, $A$ is the receiver aperture size, $\phi$ is the incidence angle, $\theta$ is the irradiance angle, and D is the inter-vehicular distance. For piece wise Lambertian model, $n$ and $\gamma$ are extracted through linear least square method. 

Linear \ac{VVLC} channel path loss model proposed in \cite{elamassie2018effect} is given by $P_{r}=P_{t}(\alpha d+\beta)$, where $P_{r}$ is received power, $P_{t}$ is transmitted power , $d$ is the inter-vehicular distance, $\alpha$ and $\beta$ are weather dependent coefficients. 

On the other hand, exponential model proposed in \cite{eldeeb2019path} is given as,

\begin{equation}
    P_{r}=P_{t}A d^{-2B}exp(-cd)
\end{equation}{}
where $d$ is intervehicular distance, c is the weather dependent extinction coefficient (i.e. $1.5x10^{-5}$ for clear weather), $A$ is geometrical loss, and $B$ is the decaying factor. 

Two term exponential model is also proposed for \ac{VVLC} path loss characterization \cite{turan2018vehicular}. The channel DC gain of two term exponential model has the following form, 
\begin{equation}
    H(0)=a_{1} e^{a_{2} D}+a_{3} e^{a_{4} D}
\end{equation}{}
where $a_{1}$, $a_{2}$, $a_{3}$ and $a_{4}$ are fitting coefficients, $D$ is the inter-vehicular distance.

Fig.~\ref{all_data} depicts normalized path loss of DS1 and path loss of DS2 as a function of inter-vehicular distance along with the best fit of two term exponential model for \ac{DS1} and all considered model fits for \ac{DS2} under all ambient light conditions.

\begin{figure}
\captionsetup[figure]{font=footnotesize,labelfont=footnotesize}
  \centering
  \begin{tabular}{@{}c@{}}
    \includegraphics[width=.8\linewidth]{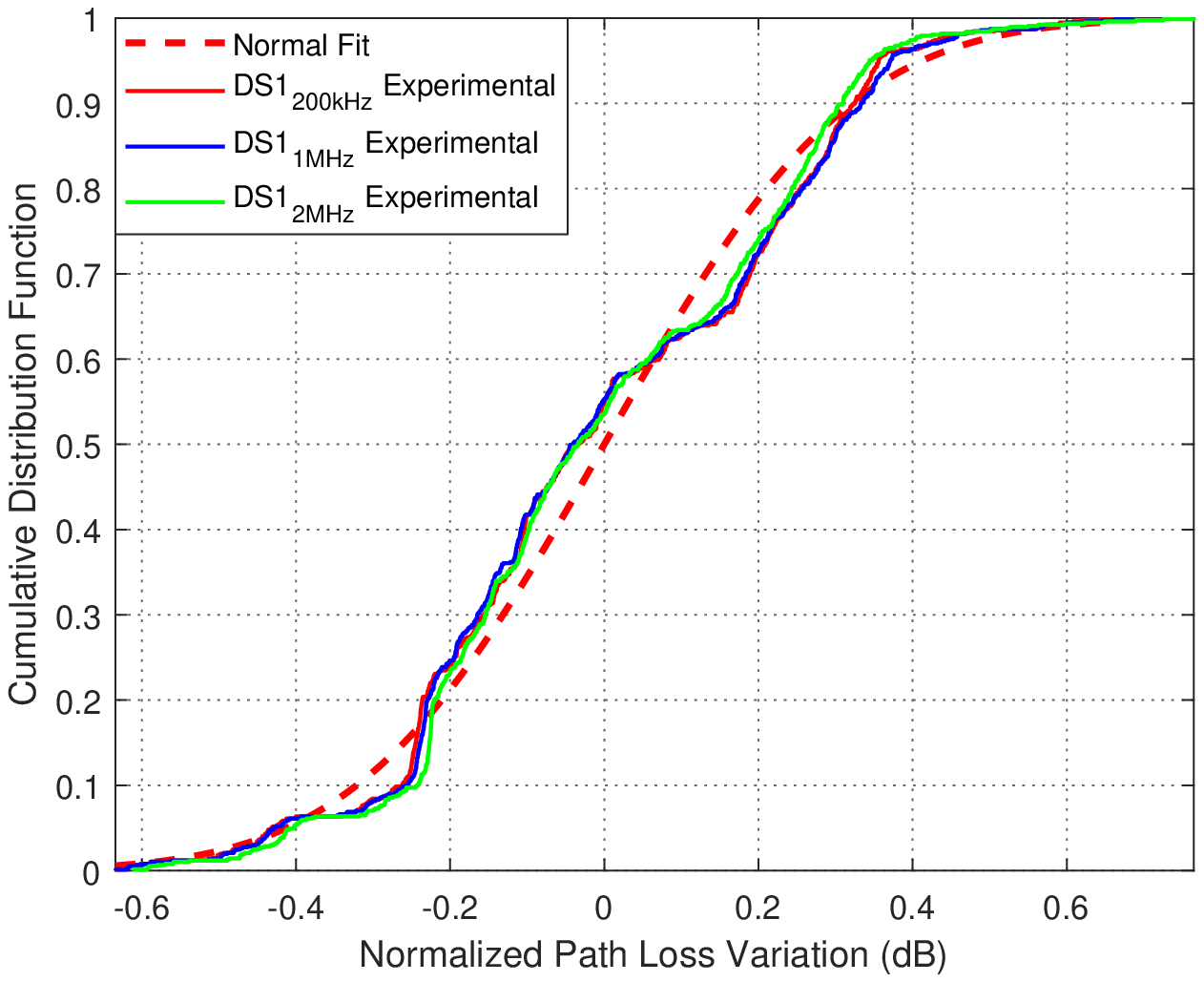} \\[\abovecaptionskip]
    \small (a)
  \end{tabular}

  \vspace{\floatsep}

  \begin{tabular}{@{}c@{}}
    \includegraphics[width=.8\linewidth]{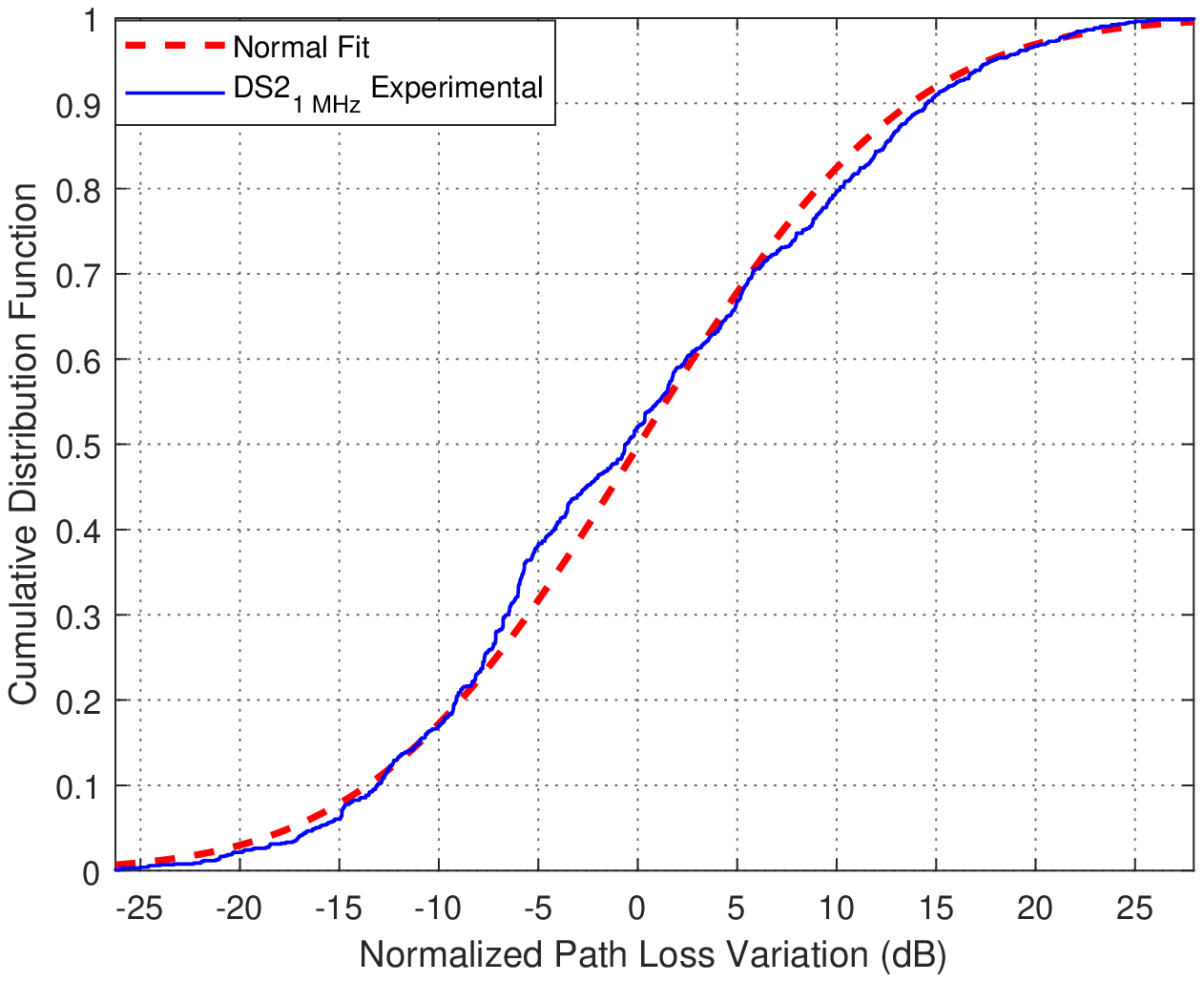} \\[\abovecaptionskip] 
    \small (b) 
  \end{tabular}

  \caption{(a) CDF of the Dataset 1 Fit Normalized Path Loss Variations (b) CDF of the Dataset 2 Fit Path Loss Variations for LoS up to 20 m }\label{CDFFits}
\end{figure}

Fig.~\ref{CDFFits} (a) shows the \ac{CDF} of experimental power variation of \ac{DS1} in decibels normalized between $-1dB$ to $1dB$. On the other hand Fig.~\ref{CDFFits} (b) depicts the path loss power variations of DS2 measurements for \ac{LoS} and same lane scenarios up to 20 m, similar conditions to DS1. The power variation denotes the difference between measured path loss and best fit through two term exponential fit in Fig.~\ref{all_data}. \ac{CDF} of power variations are compared to the frequently used normal random variable, where they are observed to deviate. Comparing power variation \ac{CDF} of two different data sets, it can be concluded that statistical generalization of \ac{VVLC} channels with respect to distance lacks accuracy due to different vehicle \ac{LED} light optics, background illumination, geometric orientation and optical turbulence.

Both static and dynamic scenario data sets, DS1, DS2,  with varying inter-vehicular distances are evaluated as benchmark for piecewise Lambertian, linear, exponential and two-term exponential models. Normalized channel path loss [$-1 dB - 1 dB$] of \ac{DS1} is considered in order to provide fair comparison due to different \ac{VNA} usage through measurements. 

\begin{table}[]
\caption{Fitted Coefficients for Existing Channel Models}
\label{fitcoeff}
\begin{adjustbox}{width=\linewidth}
\begin{tabular}{|c|c|c|c|c|c|}
\hline
\textbf{Method}                       & \textbf{Parameter} & $\mathbf{DS1_{200kHz}}$ & $\mathbf{DS1_{1MHz}}$ & $\mathbf{DS1_{2MHz}}$ & $\mathbf{DS2_{1MHz}}$ \\ \hline
\multirow{2}{*}{Piecewise Lambertian} & n                  & 45.1         & 49.04        & 20.8         & 5.883       \\ \cline{2-6} 
                                      & $\gamma$             & 1.158        & 1.119        & 0.7838       & 0.3243        \\ \hline
\multirow{2}{*}{Exponential Fitting}  & A                  & 1.148        & 1.151        & 1.182        & 0.5971        \\ \cline{2-6} 
                                      & B                  & 0.8586       & 0.8373       & 0.7536       & 0.1619        \\ \hline
\multirow{2}{*}{Linear Fitting}       & $\alpha$             & -0.06007     & -0.0585      & -0.05492     & 0.5954     \\ \cline{2-6} 
                                      & $\beta$              & 0.5061       & 0.5092       & 0.5393       & 26.21        \\ \hline
\multirow{4}{*}{Two Term Exponential}   & a                  & -0.1549      & -0.1419      & -0.09861     & 60.34       \\ \cline{2-6} 
                                      & b                  & 0.0656       & 0.06706      & 0.07597      & 0.0013     \\ \cline{2-6} 
                                      & c                  & 1.156        & 1.138        & 1.085        &-47.57        \\ \cline{2-6} 
                                      & d                  & -0.2282      & -0.2276      & -0.2191      & -0.05405       \\ \hline
\end{tabular}
\end{adjustbox}
\end{table}

Table~\ref{fitcoeff} shows the coefficients for existing \ac{VVLC} channel models extracted through least squares fit. Path loss exponent and Lambertian model order are observed to vary with respect to modulation frequency for Piecewise Lambertian model. Decaying factor (B) of exponential model is obtained close to 0.87 of \cite{eldeeb2019path} for 200 kHz and 1 MHz modulation frequencies of \ac{DS1}, where it decreases with the increasing modulation frequency.  

Table ~\ref{fitRMSE} depicts \ac{RMSE} and \ac{NoR} for \ac{DS1}, \ac{RMSE} and \ac{R^2} for \ac{DS2} fittings of considered models. \ac{RMSE} represents the standard deviation of residuals while \ac{R^2} determines how close the data is to the fitted regression line 
Considering \ac{RMSE}, \ac{NoR} and \ac{R^2} as goodness of fit metrics, two term exponential fitting outperforms the other models for both data sets and all modulation frequencies. 

\begin{figure*}[h]
\centering
\begin{minipage}[c]{.32\textwidth}
\centering
\includegraphics[trim={15 0 30 0},clip,width=\linewidth]{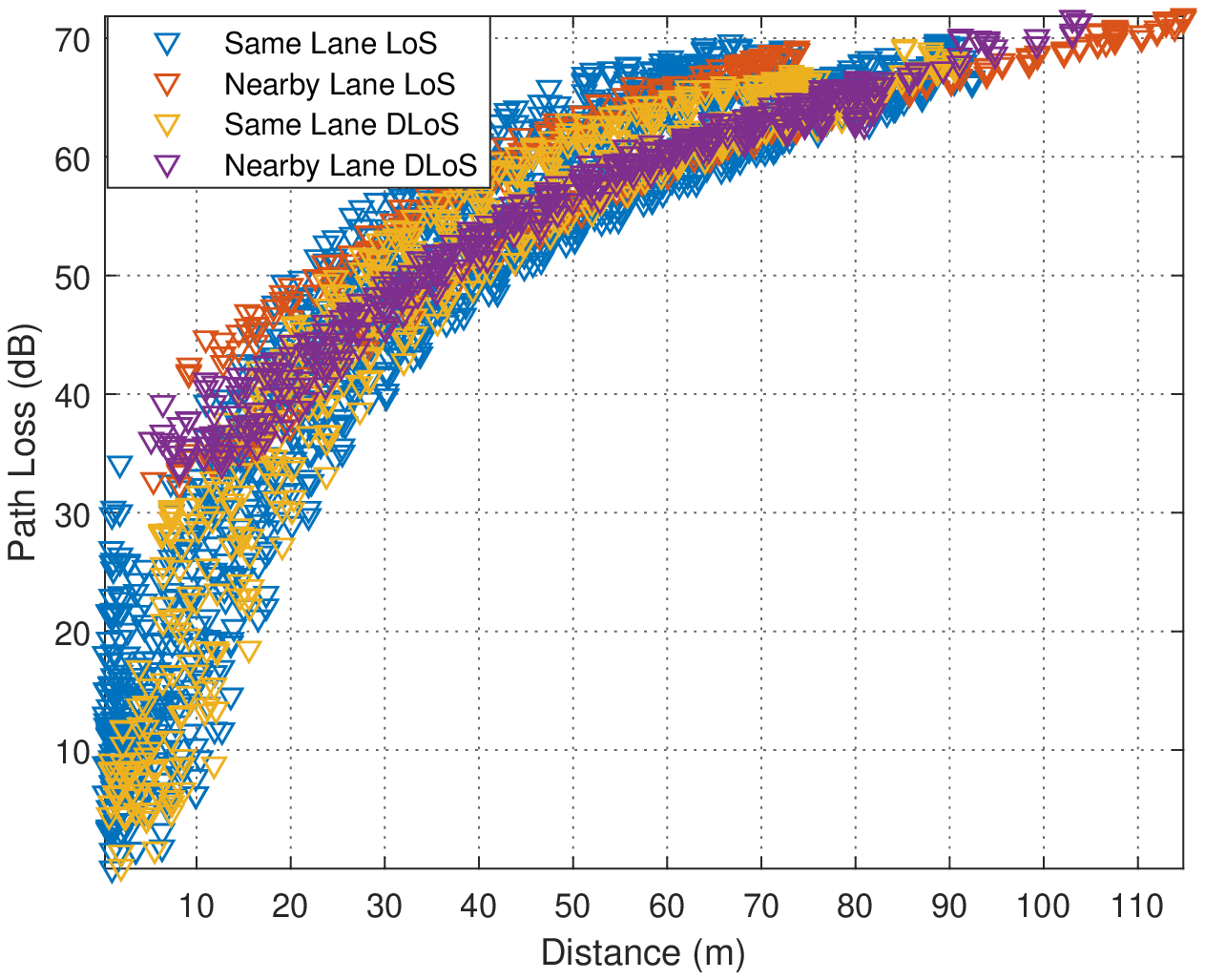}
\caption*{(a)}
\end{minipage}
\begin{minipage}[c]{.32\textwidth}
\centering
\includegraphics[trim={15 0 30 0},clip,width=\linewidth]{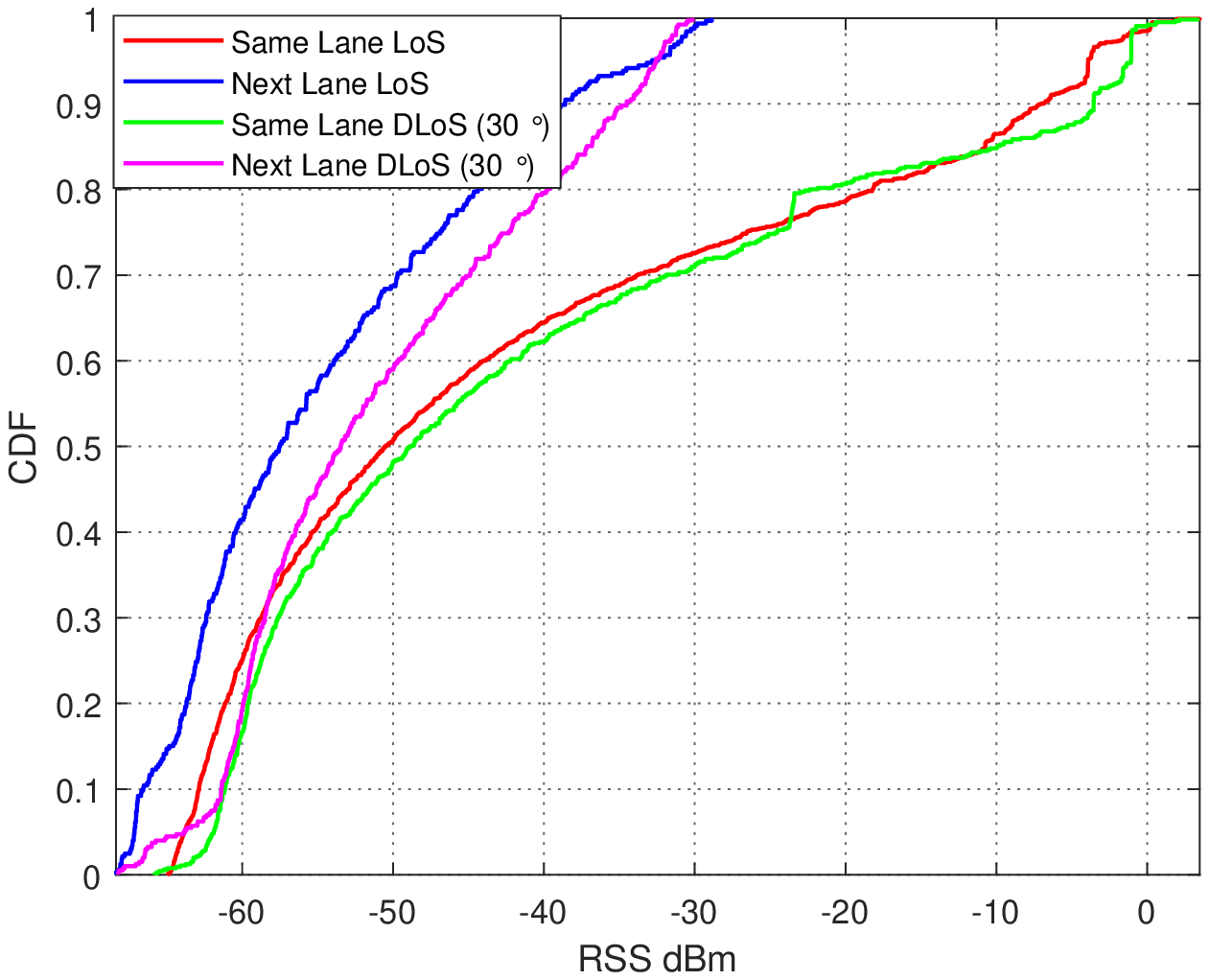}
\caption*{(b)}
\end{minipage}
\begin{minipage}[c]{.32\textwidth}
\centering
\includegraphics[trim={15 0 30 0},clip,width=\linewidth]{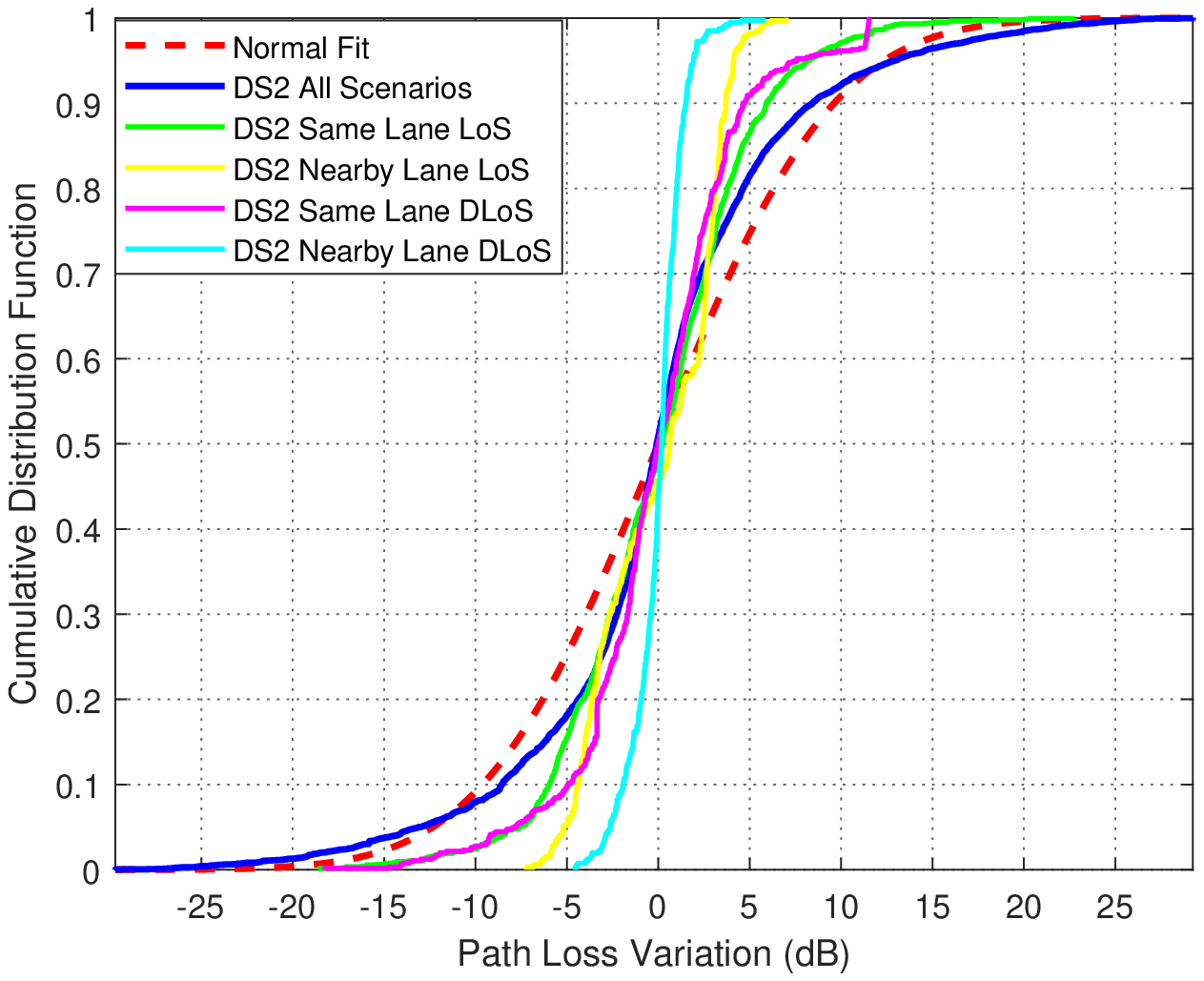} 
\caption*{(c)}
\end{minipage}
\caption{(a) Dataset 2 Distance Dependent Path Loss for All 4 Dynamic Scenarios (b) Dataset 2 CDF of RSS Values (c) Dataset 2 Two Term Exponential Fit Path Loss Variations CDF}
\label{DS3all}
\end{figure*}

Fig.~\ref{DS3all}~(a)~depicts distance dependent path loss for all dynamic scenarios of \ac{DS2}, it can be observed that most of the path loss variations occur below $20 m$ distance, where the optical beam is relatively narrow. Moreover, Fig~\ref{DS3all}~(b) reveals that road surface reflections increase \ac{DLoS} RSS for closer distances in same lane, whereas for nearby lane scenarios, both \ac{LoS} and \ac{DLoS} scenarios exhibit similar characteristics, indicating substantially weaker RSS than \ac{LoS} scenarios. Fig~\ref{DS3all}~(c) shows path loss variations of \ac{DS2} with all scenarios for two term exponential fit with the comparison of unit variance Normal distribution. All scenarios of \ac{DS2} is observed to have less power deviation than Normal distribution with unit variance. Moreover, nearby lane path loss power variations are slightly lower than same lane path loss power variations for both \ac{LoS} and \ac{DLoS} scenarios indicating a better fit for two term exponential. \ac{DS2} fitting results indicate that, for night conditions, path loss variations below $20 m$  and same lane scenarios is higher when compared to weaker illumination region of nearby lane and distances over $20 m$ . Therefore, it can be concluded that fitting based generalizations are not enough to obtain accurate \ac{VVLC} channel path loss.      

Existing channel models lack incorporating all features of ambient light, \ac{LED} frequency dependent characteristics, optical turbulence effects and reflections (i.e. road surface) with respect to receiver orientation angle effects on \ac{VVLC} channel. Thus, they provide limited generalization ability to accurately characterize \ac{VVLC} channel path loss. Therefore, \ac{VVLC} channel models generated with the consideration of ambient noise, \ac{LED} frequency dependent propagation, inter-vehicular distances and transmitter - receiver geometry (i.e. LoS , DLoS) with respect to atmospheric interactions (i.e. turbulence, scintillation) are expected to yield more accurate channel characterization.   

\begin{table}[]
\caption{Measured Channel Path Loss Goodness of Fit Metrics} 
\label{fitRMSE}
\begin{adjustbox}{width=\linewidth}
\begin{tabular}{|c|c|c|l|c|c|c|l|l|}
\hline
\multirow{2}{*}{\textbf{Method}}                                & \multicolumn{2}{c|}{{$\mathbf{DS1_{200kHz}}$}}                        & \multicolumn{2}{c|}{{$\mathbf{DS1_{1MHz}}$}}          & \multicolumn{2}{c|}{{$\mathbf{DS1_{2MHz}}$}}                       & \multicolumn{2}{c|}{{$\mathbf{DS2_{1MHz}}$}} \\ \cline{2-9} 
                                                                & \textbf{RMSE}               & \textbf{NoR}               & \textbf{RMSE} & \textbf{NoR}               & \textbf{RMSE}               & \textbf{NoR}              & \textbf{RMSE}    & \textbf{R-Square}   \\ \hline
\begin{tabular}[c]{@{}c@{}}Piecewise \\ Lambertian\end{tabular} & 0.4283                      & 11.87                      & 0.4175        & 11.58                      & 0.3841                      & 10.65                     & 7.457           & 0.8539           \\ \hline
\multicolumn{1}{|l|}{Exponential Fitting}                       & \multicolumn{1}{l|}{0.3933} & \multicolumn{1}{l|}{10.90} & 0.3804        & \multicolumn{1}{l|}{10.54} & \multicolumn{1}{l|}{0.3334} & \multicolumn{1}{l|}{9.24} & 7.459           & 0.8538           \\ \hline
Linear Fitting                                                  & 0.2671                      & 7.40                       & 0.2677        & 7.42                       & 0.2536                      & 7.03                      & 10.220          & 0.7254           \\ \hline
Two Term Exponential                                              & 0.2512                      & 6.95                       & 0.2522        & 6.98                       & 0.2394                      & 6.63                      & 7.002           & 0.8712           \\ \hline
\end{tabular}
\end{adjustbox}
\end{table}

\section{Methods and Design}
\label{Methods}


Continuous movement of vehicles leads varying inter-vehicular distances, orientation angles and different ambient light levels. 
\ac{VVLC} channel path loss and channel frequency response modeling can be classified as a regression problem, yielding relationship between path loss, ambient light,  \ac{LED} modulation frequency, transmitter-receiver distance and geometries with respect to the \ac{VVLC} front ends under consideration.  
Obtaining an analytical expression, denoting channel path loss and channel frequency response with respect to transmitter-receiver vehicle geometries, ambient light, and optical turbulence considerations is not convenient. Hence, modeling the physical parameter relationships by \glspl{NN} and Random Forest through machine learning is utilized. Therefore, channel loss data collected from different \ac{VVLC} scenarios are used to train \ac{ML} models, yielding a scenario based \ac{VVLC} channel path loss and \ac{CFR} framework. We describe the proposed \ac{ML} based channel model frameworks in Section \ref{NN}, and Section \ref{RF}.


\subsection{Neural Networks Based Channel Model Framework}
\label{NN}
Neuron is the basic component and processing unit of \glspl{NN}. Neurons produce an output vector of \glspl{NN} through multiplication of input vector $X = (x1, x2, . . . , xn)$ and its weight vectors $W = (w1,w2, . . . ,wn)$, where differentiable activation functions $f(.)$ between layers and bias $\theta$ to shift activation function are additionally employed can be generalized in the following form; 
\begin{equation}
    y= f_{oj} \left[ \sum_{j=1}^{M}w_{oj} \left( f_{ji} \left[ \sum_{i=1}^{N}w_{ji} x_{i} \right] + \theta_{j} \right) +\theta_{out}  \right] 
\end{equation}

where, $f_{oj}$ ,$w_{oj}$ and $f_{ji}$, $w_{ji}$ are activation functions and weights from neuron to output, input to neuron respectively. Minimization of the output error according to target optimization criteria (i.e. mean square error, mean absolute percentage error) is the objective of neuron models. 

Two \ac{NN} architectures, \ac{MLP} and \ac{RBF} \glspl{NN} are proposed to generalize \ac{VVLC} channel path loss and channel frequency response. 

\subsubsection{MLP}

\Ac{MLP} networks are feed-forward \glspl{NN} compromise of multiple hidden layers, involving three stages. At the first stage, input training pattern is feed-forwarded using activation functions, then associated error and weights are backpropagated through learning function. Outputs are compared with target values where the weights are readjusted to minimize the error at each iteration. Activation function is selected to be monotonically non decreasing and differentiable. For regression problems, sigmoid function in the hidden layers and linear function in the output layers are utilized. 
\Glspl{MLP}, unlike simple perceptron, are capable of classifying linearly inseparable, multivariate patterns and can solve complicated problems. 

\begin{figure}[h]
\centering
\includegraphics[width=\linewidth]{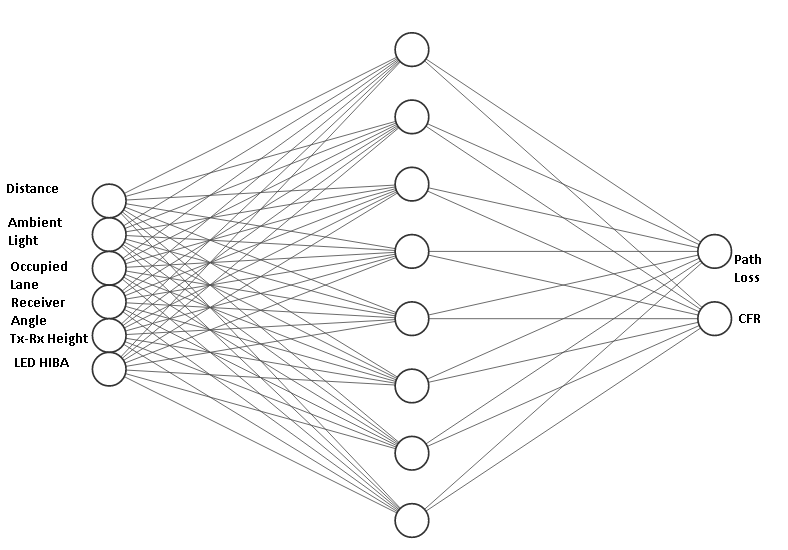}
\caption{ANN}
\label{ANN_}
\end{figure}

\subsubsection{RBF-NN}

\Glspl{RBF-NN} are three-layer feed-forward networks where the input is transformed by radially symetric basis functions at the hidden radial basis layer.
\Glspl{RBF-NN} consist of an input layer, a hidden radial basis layer with a non-linear \ac{RBF} activation function and a linear output layer. 

The number of nodes in the hidden layer depends on the complexity of the problem. However, \glspl{RBF-NN} generally require more neurons than standard \ac{MLP} networks. This is because sigmoid neurons can have outputs over a large region of the input space, while radial basis neurons only respond to relatively small regions of the input space.  \glspl{RBF-NN} perform better when many training vectors are available. Designing \ac{RBF-NN} takes less time than training a \ac{MLP} network.

The nonlinear activation function for \ac{RBF-NN} can be gaussian function, multi quadratic function, inverse multi quadratic or cauchy function. However, \cite{chen1991orthogonal} stated that the activation function selection is not crucial for performance of the \ac{RBF-NN}. Gaussian activation functions, defined by mean and standard deviation is the most common choice for \ac{RBF-NN}.

In RBF NN, the connections between the input and the hidden layers are not weighted. The inputs therefore reach the hidden layer node unchanged, and then the output of the hidden neuron is presented with the following activation function, 

\begin{equation}
\label{RBFActiv}
    G_{m}(x)=e^{-\frac{(X-V_{m})^2}{2\sigma_{m}^2}} \quad m=1,2,..,j
\end{equation}

where $X=(x_{1},x_{2}....,x_{n})$ is the input data, $V_{m}$ is the center of the $m^{\text{\tiny th}}$ neuron of the hidden layer having same dimension with $X$, $\sigma_{m}$ is the spread of the $m^{\text{\tiny th}}$ Gaussian, and $G_{m}(x)$ is the output of the $m^{\text{\tiny th}}$ Gaussian function, m denotes the total number of hidden layer nodes. Non-linear mapping of input layer  $X \rightarrow G_{m}(x) $, whereas output layer linear mapping, $G_{m}(x) \rightarrow y_{k} $, forms a linear combination of hidden layer functions with weighted sums as follows;

\begin{equation}
\label{RBFOutput}
    y_{k}=\sum_{i=1}^{m} w_{ik} G_{m}(x)  \quad k=1,2,..n
\end{equation}

where, $y_{k}$ is the output of \ac{RBF-NN}, $w_{ik}$ are the weights of linear mapping, n is the number of output layer nodes.



For the input signals closer to the centre range of the Gauss kernel, the hidden layer nodes will produce larger output. Therefore, the radial basis function network is a local approaching network and it has a superiority of fast learning speed. Spread of \ac{RBF-NN} defines the selectivity of the network, as small spread implies very selective, and many neurons are needed to obtain smooth function fit. On the other hand large spread implies less selective network output, yielding smoother function approximations. 

\Ac{RBF-NN} tend to have good interpolation properties, but not as good extrapolation properties as \glspl{MLP}. Using a given number of neurons, \ac{MLP} performs better for extrapolation purposes. On the other hand, \ac{RBF-NN} are robust to adversarial noise, due to their non-linear nature. 

\subsection{Random Forest Learning Based Channel Model}
\label{RF}
Ensemble learning, utilizing multiple individual decision trees to solve classification and regression problems, provide superior generalization performance due to its insensitive nature to variable scaling and inclusion of irrelevant variables \cite{friedman2001elements}. Random Forest is one of the prominent ensemble learning algorithms, combining estimates from multiple decision trees with random selection of features for training to yield true output through bootstrap aggregation \cite{breiman2001random}. Randomization and averaging estimates from multiple decision trees further provide robustness to noisy measurements. The maximum tree depth and the size of the ensemble determines the accuracy of the algorithm. Random Forest algorithm sort the importance of features, enabling feature dimension reduction to avoid overfitting with lower complexity models.

\section{System Model and Problem Formulation}
\label{sysmodel}

Two different model frameworks are considered yielding \ac{VVLC} channel frequency response and channel path loss predictions utilizing DS1 and DS2 respectively. 

\subsection{Channel Frequency Response Prediction Models}
With the assumption of constant electrical transmit power at swept \ac{LED} modulation frequencies, \ac{CFR} prediction of \ac{VVLC} channel is posed as a regression problem  where intra-vehicular distance, ambient light and receiver angle are inputs as $x_{i} : \{d_{i},sl_{i},\theta_{i}\}$ . The relationship between input $x_{i} \in \mathbb{R}^3$ and \ac{CFR} $y_{i} \in \mathbb{R}^{19}$ $y_{i}: \{PL_{i}^{200kHz},...,PL_{i}^{2MHz}\}$ is given by $f_{CFR} : X \rightarrow Y $ , where the objective is to estimate $f_{CFR}(x)$ using training dataset from DS1 , minimizing the \ac{CFR} estimation error for new input data that models are not trained with. 
Performance of the models are evaluated with the comparison of target values and model output $\hat{Y}$ to input test samples through \ac{RMSE}. 

\subsection{Channel Path Loss Prediction Models}

Intra-vehicular distance, \ac{LED} modulation frequency, ambient light levels, occupied lane, optical turbulence level and receiver elevation angle are inputs of the proposed models, whereas model output yields channel path loss. The VVLC channel path loss predictions can be considered as function approximation where the goal is to approximate an unknown mapping $f : X \rightarrow Y $ from a set of input parameters $X : \{d_{l},o_{l},sl_{l},lane_{l},\theta_{l}\}$ to another set of channel path losses $Y: \{PL_{l}\}$ , where $d_{l}$ denotes inter-vehicular distance, $o_{l}$ represents optical turbulence regime either low or high, $sl_{l}$ contains sun load sensor voltage values indicating ambient light, $lane_{l}$ is the occupied lane of receiver vehicle, $\theta_{l}$ is the receiver orientation angle denoting \ac{LoS} or \ac{DLoS} conditions and $PL_{l}$ , is the path loss of VVLC channel, ${d_{l},o_{l},sl_{l},lane_{l},\theta_{l}}$ and $PL_{l}$ variables are extracted from measurement data set as input-output vector pairs to train neural network and random forest during training phase. During the testing phase, the trained network is fed an input vector of $X : \{d_{l},o_{l},sl_{l},lane_{l},\theta_{l}\}$ to obtain the estimated output $\hat{Y}$. Performance of trained models are evaluated by comparing the estimate $\hat{Y}$ to the actual output $Y$ across the test data set using \ac{RMSE}. 

For path loss models 90\%, 80\%, 60\%, 30\% and 10\% of DS2 samples are utilized for training while the rest portions (10\%, 20\%, 40\%, 70\%, 90\%) are used as test set for performance evaluations.

\subsection{Data Preprocessing}
Input values of the proposed models vary in different ranges as they are different physical units. Therefore, input parameters are normalized and mapped to values  between -$1$ to $1$. Inverse conversion is executed at the output to obtain predicted values. Outlier samples captured through measurement errors are excluded from all data sets. For path loss prediction models, high variance region is observed for closer inter-vehicular distances due to the narrower beam divergence angle. Thus, k-means clustering is utilized to further label the data as high variance region and low variance region.

\subsubsection{K-means Clustering}
\label{kmeanssection}
K-means clustering iteratively partitions n observations into k non-overlapping clusters where each observation belongs to the cluster with the nearest mean. Expected maximization approach of k-means clustering assigns data points to a cluster where the sum of the squared distance between the data points and the mean of all the data points that belong to that cluster (centroid) is at the minimum. 

\begin{figure}
\centering
\includegraphics[width=\linewidth]{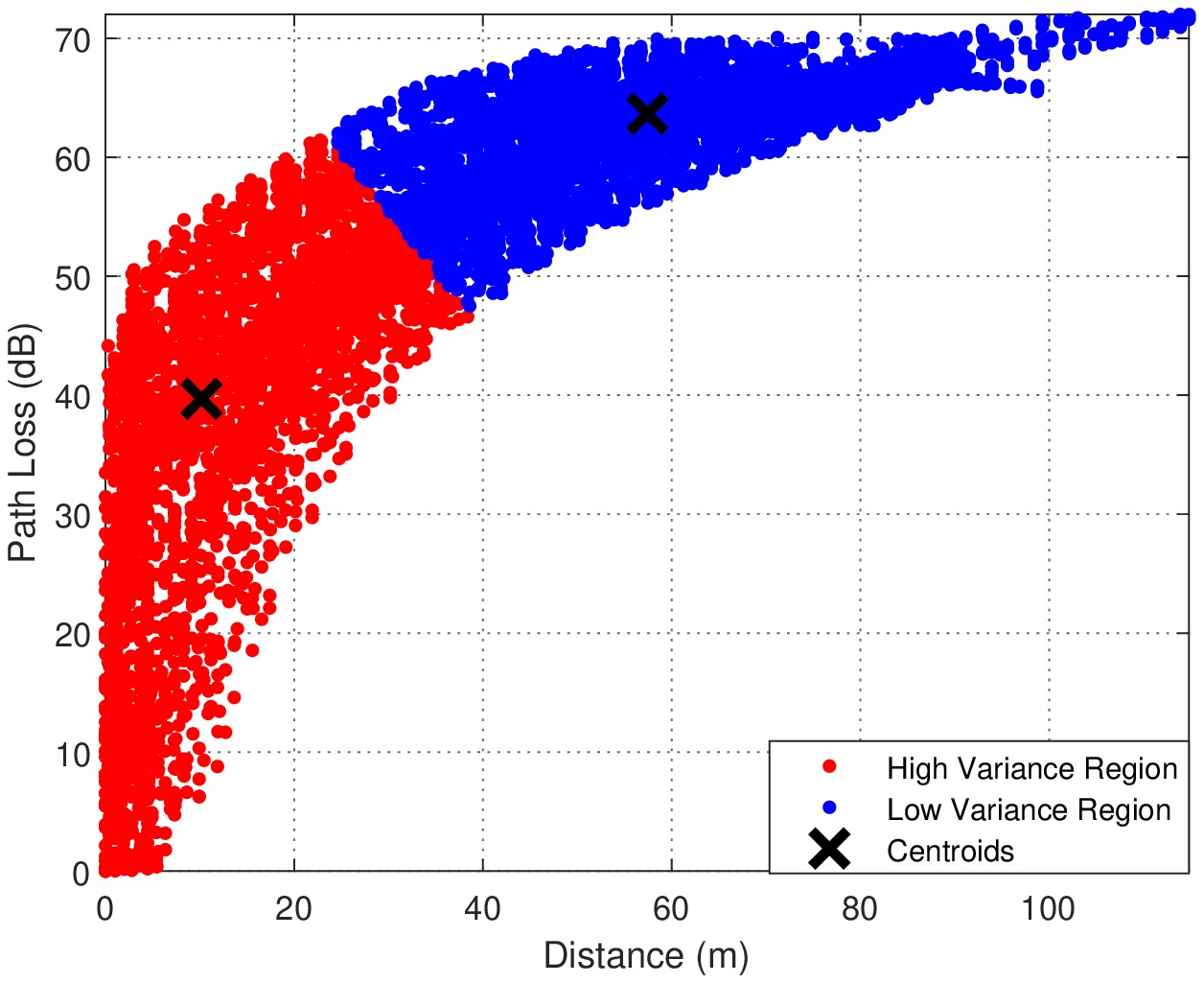}
\caption{Low Variance and High Variance Regions of DS2 from K-means clustering algorithm}
\label{kmeans}
\end{figure}

K-means clustering is conducted to define low variance and high variance regions with respect to inter-vehicular distances for channel path loss prediction models. As the initial choice of centroids can affect the output clusters, the algorithm is executed 10 times with different initializations to obtain two fair cluster partitions of DS2. 3297 samples of 7686 samples is found to be in high variance region with maximum distance of 38 m whereas 4389 samples are considered in low variance regions as depicted in Fig.\ref{kmeans}. High variance and low variance region labels are added to the training data to increase regression accuracy. 

 \subsubsection{Predictor Importance Estimation}
 Feature selection is key for both to obtain accurate results and avoid over fitting. Permutation feature importance estimation method from Random Forest is utilized to measure predictor importance \cite{breiman2001random}. For predictor importance estimation increase in the model’s prediction error after permuting the feature changes is calculated.  
 
 Table \ref{importance} denotes the normalized predictor importance values of both data sets. Distance and amplitude variance region selection appeared to be more important features than the others for \ac{VVLC} path loss. On the other hand, distance and ambient light are observed to be most important features for \ac{VVLC} \ac{CFR}. Moreover, \ac{VNA} model feature selection is perceived to have a considerable effect on the predictor performance for \ac{CFR} estimations, as \ac{VNA} drives the \ac{LED} through frequency swept electrical voltage signals and accurate calibrations can not be executed for low impedance \glspl{LED} on contrary to 50\ohm~ load. Receiver angle inclination is concluded to be the least important feature for both data sets, hence, reflections from road surface can be considered to be negligible for \ac{DLoS} when compared to \ac{LoS} transmissions.

\begin{table}[]
\caption{Normalized Importance of Input Features}
\label{importance}
\centering
\begin{adjustbox}{width=\linewidth}
\begin{tabular}{lcc}
\hline
\multicolumn{1}{c}{\textbf{Feature}} & \textbf{Importance (PL)} & \textbf{Importance (CFR)} \\ \hline
Distance                             & 0.7884                          & 0.4161                         \\ \hline
Optical Turbulence                   & 0.0579                          & -                              \\ \hline
Ambient Light                        & 0.0361                          & 0.2804                         \\ \hline
Amplitude Variance Region            & 0.0883                          & -                              \\ \hline
Occupied Lane                        & 0.0218                          & -                              \\ \hline
Receiver Inclination Angle           & 0.0075                          & 0.0661                         \\ \hline
VNA Model                            & -                               & 0.2375                         \\ \hline
\end{tabular}
\end{adjustbox}
\end{table}


\subsection{Hyperparameter Selection}
\ac{ML} model parameters, that can be selected before training process are known as hyperparameters. Grid search, random search and Bayesian optimization methods are commonly utilized for hyperparameter optimization. Hyperparameter optimization plays an important role for \ac{ML} models to obtain accurate prediction results in reduced training time, while ensuring simultaneous convergence, hence avoiding over fitting. 

In this work, grid search is employed to find the optimal combination of hyperparameters through search of all possible points in the given range. For \ac{MLP} the following parameter intervals are considered for both \ac{CFR} and path loss models : number of neurons for 2 hidden layers between 1 to 50 and the number of maximum validation failures between 3 to 10, minimum performance gradient between 1e-3 to 1e-8 with 1e-5 intervals.  Considering \ac{RBF-NN}, selection of spread is important to obtain smoother function. Therefore, spread values are evaluated between 0.2 to 10 for both \ac{CFR} and path loss estimation models. For path loss predictions with Random Forest, the number of decision trees between 150 and 300 and the maximum tree depths between 16 and 1024 are evaluated. 

The optimum hyperparameters yielding best predictions are obtained as follows. For \ac{CFR} prediction MLP model, 30 neurons at first hidden layer, 20 neurons at second hidden layer, whereas for 20 neurons at first hidden layer and 10 neurons at second hidden layer for \ac{CFR}. Maximum validation failure of 5, minimum performance gradient of 1e-6 are set for both \ac{MLP} networks. \ac{RBF-NN} spread factor for path loss is 0.2 and 0.7 for \ac{CFR}. For Random Forest, the depth of trees and the number of decision trees are finally chosen as 253 and 710, respectively for path loss predictor. 

\subsection{Model Implementation}

We developed the proposed models using MATLAB software running on Dell T5610 workstation. The workstation is equipped with NVIDIA Quadro K2000 graphics card and 12 CPU cores, enabling parallel training on GPU and CPU multi-cores.

\subsubsection{\ac{MLP} Framework for VVLC Channel Path Loss Predictions}

Data sets of \ac{MLP} networks are splitted as 60\% training, 20\% testing and 20\% validation, then the input features were scaled between $-1$ to $1$.  
Five fold cross validation scheme is employed to determine best model for \ac{MLP}. 

Our network is multi-layer perceptron (MLP) type feed forward architecture. It is based on a supervised training using scaled conjugate gradient back propagation. We use hyperbolic tangent sigmoid function (Tansig) in the hidden layers and linear function (Purline) in the output layer.

\ac{MLP} networks for both \ac{CFR} and path loss predictions are modeled with a perceptron of three layers, two hidden layers and one output layer. The number of neurons per layer is varied to improve performance. 

\subsubsection{\ac{RBF} Framework for VVLC Channel Path Loss Predictions}

The RBF framework created for VVLC channel path loss predictions has the Gaussian function as activation function in its hidden layer, and linear
function in its output layer.  Training of the hidden layer involves the determination of the  radial basis functions by specifying appropriate  $\sigma_{m}$ values of (\ref{RBFActiv}). This parameter depend only on the input data and are independent of the outputs, yielding unsupervised learning. On the other hand, output layer is trained by a supervised learning method, where the synaptic weights are updated in proportion to the difference between the network and target output.  The input data is scaled between -1 and 1, where 70\% of data sets is used for training and 30 \% of the all samples are used for testing \glspl{RBF-NN}.


The spread of the Gaussian for \ac{CFR} estimator \ac{RBF-NN}, and path loss estimator \ac{RBF-NN}, is defined as 0.2 and 0.7 respectively. The training parameter goal for both networks is set to 1e-1. 

\subsubsection{Random Forest}

For Random Forest algorithm, decision tree depth denotes the number of splits made on the independent variables. The number of decision trees that their outputs are averaged over gives the size of the ensemble. Too deep trees with small ensemble size lead to detailed models with overfitting, whereas too shallow trees might yield overly simplified models that can not fit data accurately. Generally, increasing the ensemble size makes the model more robust. However, the improvement decreases after certain number of added decision trees, where the cost in computation time for learning should be considered. 
Random forest used 60\% data for training and the rest is used for validation.


Random  Forest algorithm for path loss predictions is designed with the following parameters, 
number of decision trees of 253,  maximum number of splits of 710.  \Ac{MSE} is used as the performance  criteria through 10 fold cross validations.

\section{Performance Evaluation}

\begin{table}[]
\caption{MLP Based Path Loss Estimation Framework Parameters and Prediction Performance} 
\label{MLPPL}
\begin{adjustbox}{width=\linewidth}
\begin{tabular}{|c|c|c|c|c|}
\hline
\textbf{Training Dataset (DS2)} & \textbf{\begin{tabular}[c]{@{}c@{}}Number of Neurons\\ (Layer 1 -2 )\end{tabular}} & \textbf{\begin{tabular}[c]{@{}c@{}}MAE (dB) \end{tabular}} & \textbf{\begin{tabular}[c]{@{}c@{}}RMSE (dB)\end{tabular}} & \textbf{\begin{tabular}[c]{@{}c@{}}R-Correlation Coefficient\end{tabular}} \\ \hline
10 \%                  & 35-20                                                                                   & 2.7735                                                                                                      & 4.8249                                                                    & 0.9642                                                                   \\ \hline
30 \%                  & 33-15                                                                              & 2.2874                                                                                         & 3.9881                                                             &0.9742                                                            \\ \hline
60 \%                  & 31-10                                                                              & 2.1856                                                                                         & 3.9502                                                             & 0.9755                                                                   \\ \hline
80 \%                  & 39-29                                                                             & 2.2676                                                                                        & 4.0885                                                         & 0.9746                                                                    \\ \hline
90 \%                  & 37-20                                                                             & 2.2702                                                                                         & 4.2637                                                        & 0.9700                                                                   \\ \hline
\end{tabular}
\end{adjustbox}
\end{table}

\label{Performance}

\begin{table}[]
\caption{RBF-NN Based Path Loss Framework Parameters and Prediction Performance} 
\label{RBFNNPERF}
\begin{adjustbox}{width=\linewidth}
\begin{tabular}{|c|c|c|c|c|}
\hline
\multicolumn{1}{|l|}{\textbf{Training Dataset (DS2)}} & \multicolumn{1}{l|}{\textbf{Number of Neurons}} & \multicolumn{1}{l|}{\textbf{MAE (dB)}} & \multicolumn{1}{l|}{\textbf{RMSE (dB)}} & \multicolumn{1}{c|}{\textbf{\begin{tabular}[c]{@{}c@{}}R-Correlation Coefficient\end{tabular}}} \\ \hline
10 \%                                        & 289                                             & 1.8938                                           & 3.6063                             &0.9794                                                                                      \\ \hline
30 \%                                        & 289                                             & 1.8682                                           & 3.5037                              &0.9807                                                                                      \\ \hline
60 \%                                        & 551                                             & 1.8854                                           & 3.5305                             &0.9799                                                                                      \\ \hline
80 \%                                     & 551                                             & 1.8606                                           & 3.5426                             &0.9799                                                                                      \\ \hline
90 \%                                        & 551                                             & 2.2041                                          & 4.2858                             &0.9722                                                                                      \\ \hline
\end{tabular}
\end{adjustbox}
\end{table}

We evaluated the performance of \ac{MLP}, \ac{RBF-NN} and Random Forest methods for \ac{VVLC} channel path loss predictions, whereas \ac{MLP} and \ac{RBF-NN} models are considered for \ac{CFR} estimations. \ac{RMSE} and \ac{MAE} metrics used to evaluate model performances are given as,

\begin{equation}
\label{metrics}
\begin{aligned}
    MAE=\frac{1}{n} \sum_{i=1}^{n} |T_{i}-O_{i}| \\
    RMSE=\sqrt{\frac{1}{n} \sum_{i=1}^{n} (T_{i}-O_{i})^2}
    \end{aligned}
\end{equation}

where, $T_{i}$ is the target value of i th test sample, $O_{i}$ is the model output value of the i th sample from test set, n is the total number of test set samples.

For path loss prediction models, various portions of measurement samples are randomly selected to train the network, while the rest is utilized to test the networks. Table \ref{MLPPL} depicts the performance of \ac{MLP} path loss models with respect to trained data, whereas Table \ref{RBFNNPERF} depicts the \ac{RBF-NN} path loss model performance results. Both networks with various training sample sizes are observed to yield better prediction performance when compared to fitting based VVLC models. Moreover, the performance of \ac{MLP} based path loss prediction model decreases with the increasing number of training samples, indicating overfitting. Therefore, 60\% of samples are considered to train both \ac{MLP} and \ac{RBF-NN} networks, while the rest 40\% is utilized to test the network for optimum performance results. \ac{RBF-NN} outperforms \ac{MLP} 0.42 dB \ac{RMSE} and 0.3 dB \ac{MAE} for path loss predictions, considering same training and test samples for both networks. \ac{RBF-NN} requires more neurons than \ac{MLP} for similar prediction performance, whereas the training time of \ac{RBF-NN} is substantially lower than the training time of \ac{MLP} (i.e. 20 mins for \ac{RBF-NN}, 13 hours for \ac{MLP})

\captionsetup[figure]{font=footnotesize,labelfont=footnotesize} 
\begin{figure}[h]
    \centering
  \subfloat[\label{mlphist}]{%
       \includegraphics[width=0.5\linewidth]{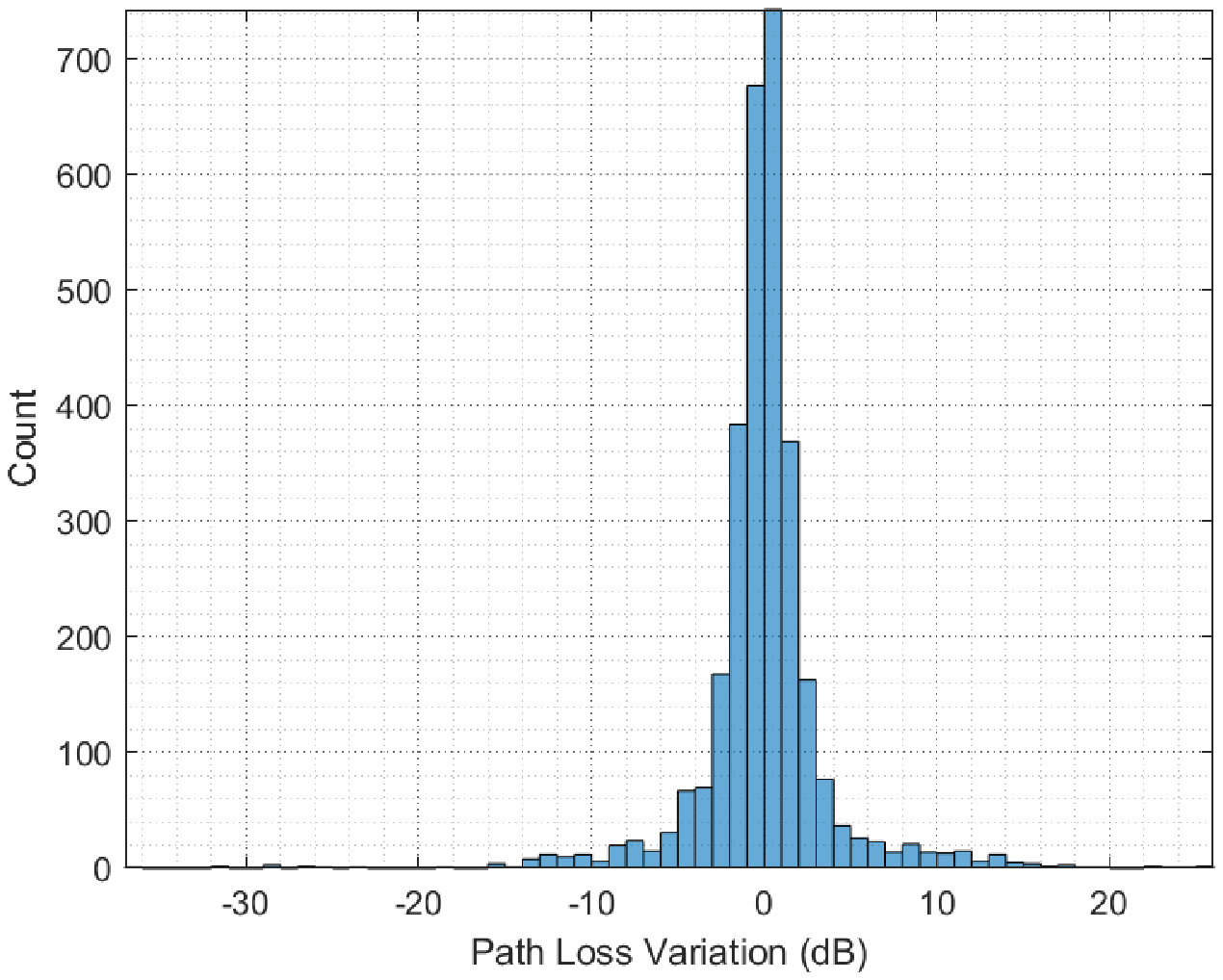}}
    \hfill
  \subfloat[\label{rbfhis}]{%
        \includegraphics[width=0.5\linewidth]{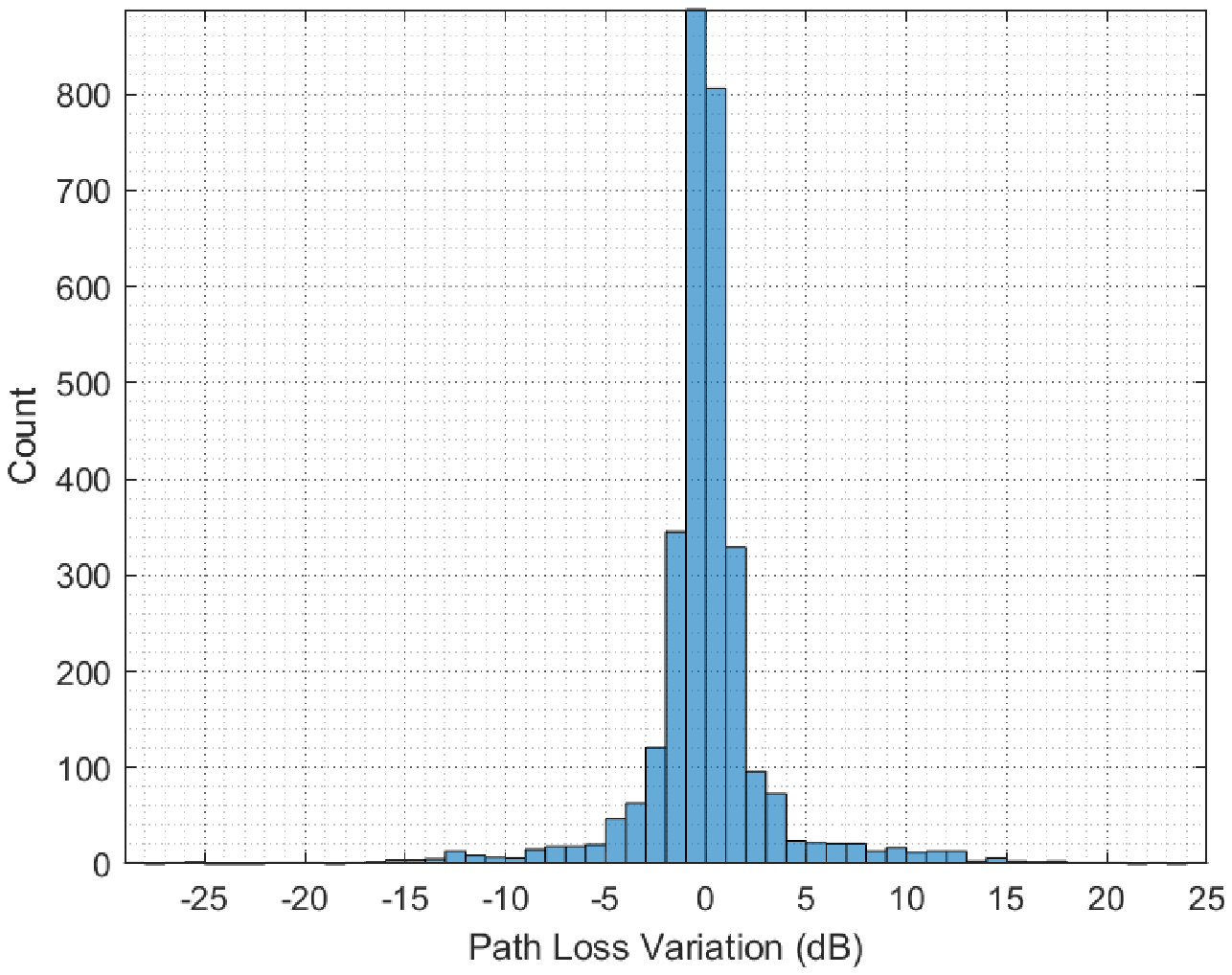}}
    \\
  \caption{(a) Prediction Error Distribution of MLP Based Path Loss Model  (b) Prediction Error Distribution of RBF-NN Based Path Loss Model}
  \label{PE}
\end{figure}

\begin{table}[h]
\caption{Best Path Loss Models with Optimal Hyperparameters}
\label{BESTPL}
\begin{adjustbox}{width=\linewidth}
\begin{tabular}{cccc}
\hline
\textbf{Algorithm} & \textbf{Optimal Hyperparameters}  & \textbf{RMSE (dB)} & \textbf{MAE (dB)} \\ \hline
Random Forest      & Number of Estimators 253 , Maximum Depth 710 & 3.8107             & 2.4541            \\ \hline
MLP                & 35-10 2 layer network, tansig activation function                                   & 3.9502                  & 2.1856                  \\ \hline
RBF-NN             & Spread Factor 0.4,  NN Size 551                                 & 3.5305                   & 1.8854                  \\ \hline
\end{tabular}
\end{adjustbox}
\end{table}

Table \ref{BESTPL} shows the selected optimal hyperparameters along with \ac{MAE} and \ac{RMSE} values for each method. \ac{RBF-NN} outperforms both methods, whereas Random Forest yields lower \ac{RMSE} but higher \ac{MAE} when compared to \ac{MLP}.

Table \ref{BESTCFR} depicts the performance results of \ac{MLP} and \ac{RBF-NN} to model \ac{CFR} with selected hyperparameters. \ac{RBF-NN} is observed to outperform \ac{MLP} by  0.18 dB RMSE whereas \ac{MLP} \ac{MAE} is 0.97 dB is less than \ac{RBF-NN} for \ac{CFR} estimations. Therefore, \ac{CFR} of a \ac{VVLC} channel can be predicted with similar performance using both models with respect to ambient light, inter-vehicular distance and receiver inclination angle. 

\begin{table}[h]
\caption{Best CFR Models with Optimal Hyperparameters}
\label{BESTCFR}
\begin{adjustbox}{width=\linewidth}
\begin{tabular}{cccc}
\hline
\textbf{Algorithm} & \textbf{Optimal Hyperparameters}  & \textbf{RMSE (dB)} & \textbf{MAE (dB)} \\ \hline
MLP                & 27-15 2 layer network, tansig activation function                                   & 3.7801                  & 2.6173                \\ \hline
RBF-NN             & Spread Factor 0.2 , NN Size 55                               & 3.6043                  &     3.5821                 \\ \hline
\end{tabular}
\end{adjustbox}
\end{table}

\begin{table}[]
\caption{CFR Estimation Model Performance with Different Training Data Size} 
\label{CFRCOMP}
\centering
\begin{tabular}{cccc}
\hline
\textbf{Training Data Size} & \textbf{Model} & \textbf{MAE (dB)} & \textbf{RMSE (dB)} \\ \hline
30\%                   & RBF-NN           & 4.1884            & 4.2120             \\ \hline
70\%                   & RBF-NN           & 3.5821            & 3.6043             \\ \hline
30\%                   & MLP (28-35)             & 2.6428            & 3.8027                   \\ \hline
70\%                   & MLP (27-15)            & 2.6173            & 3.7801                   \\ \hline
\end{tabular}
\end{table}

\ac{MLP} model for \ac{CFR} predictions is observed to perform similar with reduced training data set. However, \ac{RBF-NN} prediction performance degrades 0.6 dB when the training data size reduces from 70 \% of all samples to 30 \% as depicted in Table~\ref{CFRCOMP}.

\section{Conclusion} \label{conclusion}

This work introduces a novel approach than traditional methods in modeling the VVLC channel loss and \ac{CFR} on a practical road environment, based on \ac{ML} techniques. The validation results based on experimental measurements demonstrate the efficiency of the proposed frameworks to predict or generate \ac{VVLC} channel path loss with respect to relevant input parameters. 

Revealing the importance of the features affecting \ac{VVLC} channel performance through ensemble learning, accurate channel loss and \ac{CFR} predictions can be obtained. Moreover, data acquisition for channel modelling can be executed in a systematic manner, as more concentration will be given to relatively important features. However, the importance of the features can be experimental setup dependent. Increased variance in the captured data of the feature increases its importance. For example, with wide \ac{FoV} optical receiver angular orientation may be less important when compared to narrower \ac{FoV} receiver. Therefore, experimental setup plays an important role for the feature importance selection and generalization ability of the \ac{ML} model. 

For \ac{VVLC} channels, \ac{ML} techniques are demonstrated to yield better generalization than fitting based models, even for the reduced amount of training data. Therefore, the main drawback of big data acquisition to train \ac{ML} models is not the case for \ac{VVLC} channel modelling through \ac{ML} methods. Prediction accuracy obtained through the proposed \ac{ML} methods can be further increased through refinement of the models with the evaluation of various algorithms and hyper parameter optimization. 

Proposed \ac{ML} models built and validated through real world measurements, enable new scenario based data set generation. As they are not constrained with probabilistic distributions, analytical expressions, and assumptions, parameters extracted from generated data sets and ML based channel models will be closer to field measurements.

\bibliographystyle{IEEEtran}
\bibliography{references}

\end{document}